\documentclass[preprint,prd,tightenlines,nofootinbib]{revtex4}
\usepackage{graphicx}
\usepackage{dcolumn}
\def \MSbar {\vbox{\hrule\kern 1pt\hbox{\rm MS}}}
\begin{document}
\title{Next-to-leading order numerical calculations in Coulomb gauge}
\author{ Michael Kr\"amer}
\affiliation{Department of Physics and Astronomy, 
University of Edinburgh, Edinburgh EH9 3JZ, Scotland}
\author{ Davison E.\ Soper}
\affiliation{Institute of Theoretical Science, 
University of Oregon, Eugene, OR 97403 USA}
\date{9 April 2002}

\begin{abstract}
Calculations of observables in quantum chromodynamics can be performed
using a method in which all of the integrations, including integrations
over virtual loop momenta, are performed numerically. We use the
flexibility inherent in this method in order to perform next-to-leading
order calculations for event shape variables in electron-positron
annihilation in Coulomb gauge. The use of Coulomb gauge provides the
potential to go beyond a purely order $\alpha_s^2$ calculation by
including, for instance, renormalon or parton showering effects. We
expect that the approximations needed to include such effects at all
orders in $\alpha_s$ will be simplest in a gauge in which unphysically
polarized gluons do not propagate over long distances.
\end{abstract}

\pacs{}
\maketitle

\section{Introduction}
\label{sec:introduction}

QCD calculations at next-to-leading order can be done in a style in
which all of the integrations over loop three-momenta are done
numerically. In this method, in particular, the integrations for virtual
loop diagrams are performed numerically instead of analytically. The
method works well in Feynman gauge
\cite{beowulfPRL,beowulfPRD,beowulfrho} when applied to three-jet-like
observables in electron-positron annihilation ({\it eg.}  the three jet
cross section).

Feynman gauge is the simplest gauge from a calculational point of view.
However, it is unphysical in that unphysical gluon polarizations
propagate into the final state. The contributions from unphysical
polarizations cancel when one sums over graphs. However, we have in
mind applications in which one wants to go beyond a pure
next-to-leading order calculation by incorporating, in an approximate
way, some effects at all orders in $\alpha_s$. (We have in mind, for
instance, renormalon and parton showering effects.) For such
applications, one must approximate, and the presence of unphysical
degrees of freedom propagating over long distances makes it difficult
to see what approximations to apply. The remedy is simple: do the
calculation in a physical gauge, such as Coulomb gauge.

In this paper, we develop the apparatus needed for applying the
numerical integration method in Coulomb gauge.\footnote{ We choose
Coulomb gauge over space-like axial gauge because we have chosen to
treat cross sections in electron-positron annihilation, which has a
natural symmetry under rotations in the electron-positron c.m.\ frame.
The choice of Coulomb gauge maintains this symmetry. Time-like axial
gauge might have been a good choice, but this choice complicates the
structure of amplitudes as a function of the energy in a virtual loop.}
For the most part, this is straightforward: one should simply replace
the Feynman gauge Feynman rules by the Coulomb gauge Feynman rules.
However, two point functions and three point functions need a special
treatment (in any gauge). Three point one loop virtual subgraphs need
a special treatment because they are ultraviolet divergent. The
modified minimal subtraction (\MSbar) prescription to calculate in
$3-2\epsilon$ space dimensions and remove poles is not useful for
numerical integrations. Thus one must convert the \MSbar\ subtraction
to an equivalent subtraction defined in exactly 3 dimensions. (Four
point subgraphs with all gluon legs need renormalization too, but such
subgraphs do not occur in the applications that we have in mind, so we
omit consideration of them.) The two point one loop virtual subgraphs
need a special treatment because they need renormalization. They also
need a special treatment for another reason. Let $\Sigma(q)$ be the one
loop quark self-energy function. Then when the self-energy attaches to
a quark line that enters the final state, we need $\rlap{/}q\Sigma(q)
\rlap{/}q/q^2$ evaluated at $q^2 = 0$. We need to express $\Sigma(q)$
as a numerical integral, but we need to do it in such a way that the
{\it integrand} for $\rlap{/}q\Sigma(q) \rlap{/}q/q^2$ is finite at
$q^2 = 0$.  The integral will then have a logarithmic infrared
divergence, but this infrared divergence will be cancelled by a
corresponding divergence in the graph with a cut self-energy diagram.
The same consideration applies to the gluon propagator.

This paper will also serve to document the methods needed to treat
two-point subgraphs and three-point virtual subgraphs even in Feynman
gauge. These methods were discussed briefly in \cite{beowulfPRL} but the
details were left to unpublished notes \cite{beowulfnotes} that
accompany the associated computer code \cite{beowulfcode}.

We have implemented the methods described in this paper in a computer
program \cite{beowulfcode}. The code is based on the Feynman gauge code
described in \cite{beowulfPRL,beowulfPRD,beowulfrho}. In its default
mode, the program acts as a next-to-leading order Monte Carlo event
generator, generating three and four parton final states with
corresponding weights. Suppose, for example, one wishes to calculate
the expectation value of $(1-t)^n$, where $t$ is the thrust of each
event and $n$ is fixed. To do this, a separate subroutine calculates
$(1-t)^n$ for each event, multiplies by the corresponding weight, and
averages over events. As in other programs of this type, the weights
can be positive or negative. The user can specify either Feynman or
Coulomb gauge for the calculation. Although the graph-by-graph
contributions are very much gauge dependent, the net results are the
same for the two gauges.

The plan of this paper is as follows. We begin in Sec.~\ref{sec:style}
with a brief review of the main structure of a next-to-leading order
calculation by numerical integration. Then in Sec.~\ref{sec:elliptical}
we explain the momentum space coordinates used in the analysis of two
point functions. In Secs.~\ref{sec:cutgluonstructure},
\ref{sec:realglue} and \ref{sec:virtualglue}, we analyze the one loop
gluon self-energy in Coulomb gauge. In
Secs.~\ref{sec:cutquarkstructure}, \ref{sec:realquarks}, and
\ref{sec:virtualquarks} we turn to the one loop quark self-energy. In
Sec.~\ref{sec:3pointrenormalization}, we examine the renormalization of
three point functions in Coulomb gauge. Finally, in
Sec.~\ref{sec:results}, we present some results from the computer
program that implements the formulas from this paper. Appendix
\ref{sec:appendix} contains formulas for Feynman gauge that correspond
to the Coulomb gauge formulas of the main body of the text.

\section{Calculations by numerical integration}
\label{sec:style}

The idea of this paper is to make the numerical method for
next-to-leading order QCD calculations work in Coulomb gauge. Before
beginning this task, we need to outline the numerical method itself,
which works in any gauge. We present a sketch only since the
details can be found in Refs.~\cite{beowulfPRL,beowulfPRD,beowulfrho}.
Besides the matter of the gauge choice, there is one difference between
the computer algorithms presented here and those of
Refs.~\cite{beowulfPRL,beowulfPRD,beowulfrho}: here we wish to
calculate the complete observable at next-to-leading order, that is the
sum of the contributions proportional to $(\alpha_s/\pi)^1$ and 
$(\alpha_s/\pi)^2$, rather than just the coefficient of
$(\alpha_s/\pi)^2$. This is a rather trivial change, but it affects the
way the problem is set up below.

We wish to calculate an observable ${\cal I}$ with the following
structure
\begin{equation}
{\cal I} = { 1 \over \sigma_0\left(\sqrt S\, \right)}
\sum_n { 1 \over n!}\,\int d\vec P_1\cdots d\vec P_n\
{ d\sigma \over d\vec P_1\cdots d\vec P_n}\,
{\cal S}_n(\vec P_1,\dots ,\vec P_n).
\end{equation}
We work in the $e^+ e^-$ c.m.\ frame and $\sqrt S$ is the c.m.\ 
energy.\footnote{We also average over the direction of the beam axis
relative to the $z$ axis of our coordinate system. Thus we can
calculate typical event shape variables like the cross section to make
three jets, but not correlations between a jet direction and the beam
axis. However, nothing in the general methods used here prevents one
from removing this simplification.} The observable ${\cal I}$ has been
normalized by dividing by the Born level cross section $\sigma_0$ for
$e^+ + e^ - \to {\it hadrons}$. The quantity ${ d\sigma /[ d\vec
P_1\cdots d\vec P_n]}$ is the cross section to make $n$
massless partons.  In defining this cross section, we treat all $n$
partons as identical and thus divide by $n!$. The
functions ${\cal S}_n$ are the measurement functions that define the
observable \cite{KS}. They are symmetric under interchange of any of
their variables and have the property of infrared safety:
\begin{equation}
{\cal S}_{n+1}\!\left(\vec P_1,\dots, \lambda\vec P_n,
(1-\lambda)\vec P_n\right) 
= {\cal S}_{n}\!\left(\vec P_1,\dots,\vec P_n\right)
\label{IRsafety}
\end{equation}
for $0 \le \lambda < 1$. Typically the functions ${\cal S}_n$ are
dimensionless, but we do not assume that here. We are concerned with
three-jet-like quantities, which means that ${\cal S}_2 = 0$, so that
the smallest value of $n$  that contributes is $n = 3$. We work in
next-to-leading order perturbation theory, so that there are at most
four partons in the final state. Thus the sum over $n$ runs over $n =
3$ and $n = 4$.

The parton level cross sections contain delta functions, which we make
explicit by writing
\begin{equation}
{ d\sigma \over d\vec P_1\cdots d\vec P_n} =
\delta\left(\sum\vec P_i\right)\,
\delta\left(\sum |\vec P_i| - \sqrt S\right)
F_{\!n}\!\left(\vec P_1,\dots,\vec P_n,\sqrt S,
\alpha_s\!\left(C \sqrt S\right)\right).
\end{equation}
The function $F_{\!n}$ depends on the momenta and on $\alpha_s$, which
we evaluate at a scale $\mu = C\sqrt S$, where $C$ is a dimensionless
parameter of order 1. The order $\alpha_s^2$ contributions contain
logarithms of $\mu$ and thus of $\sqrt S$, so we have indicated a
separate dependence on $\sqrt S$. The dependence of $F$ on the
dimensionless parameter $C$ is left implicit. Thus we write ${\cal I}$
as
\begin{eqnarray}
{\cal I} &=& 
{ 1 \over \sigma_0\left(\sqrt S\, \right)}
\sum_n { 1 \over n!}\,\int d\vec P_1\cdots d\vec P_n\
\delta\left(\sum\vec P_i\right)\,
\delta\left(\sum |\vec P_i| - \sqrt S\right)
\nonumber\\
&&\times 
F_{\!n}\!\left(\vec P_1,\dots,\vec P_n,\sqrt S,
\alpha_s\!\left(C \sqrt S\right)\right)
{\cal S}_n(\vec P_1,\dots ,\vec P_n).
\end{eqnarray}

The energy conserving delta function would create problems in a
numerical integration, so we get rid of it by the following strategy.
We introduce a factor 1 written as
\begin{equation}
1 = \sqrt S\int_0^\infty \! d t \ h\left(t \sqrt S\right),
\label{hdef}
\end{equation}
where $t$ has the dimensions of time and $h$ is any convenient
smooth function whose integral is 1. We change integration variables in
the momentum integrals to dimensionless variables
\begin{equation}
\vec p_i = t \vec P_i\ .
\end{equation}
Dimensional analysis gives
\begin{eqnarray}
F_{\!n}\!\left(\vec p_1/t,\dots,\vec p_n/t,\sqrt S,
\alpha_s(C \sqrt S)\right) &=&
t^{3n - 2}
F\!\left(\vec p_1,\dots,\vec p_n,t\sqrt S,
\alpha_s\!\left(C \sqrt S\right)\right),
\nonumber\\
\sigma_0\left(\sqrt S\, \right)
&=&
t^2\,\sigma_0\left(t\sqrt S\, \right).
\end{eqnarray}
Thus
\begin{eqnarray}
{\cal I} &=& 
{ 1 \over \sigma_0\left(t\sqrt S\, \right)}
\sum_n { 1 \over n!}\,\int d\vec p_1\cdots d\vec p_n\
\int\! dt\, h\!\left(t\sqrt S\right)\,
\delta\left(\sum\vec p_i\right)\,
\delta\left(\sum |\vec p_i|/\sqrt S - t\right)
\nonumber\\
&&\times 
F_{\!n}\!\left(\vec p_1,\dots,\vec p_n,t\sqrt S,
\alpha_s\!\left(C \sqrt S\right)\right)
{\cal S}_n(\vec p_1/t,\dots ,\vec p_n/t).
\end{eqnarray}
Now we can use the integration over $t$ to eliminate the
energy-conserving delta function. Denoting
\begin{equation}
\sqrt s \equiv \sum |\vec p_i|
\end{equation}
we have
\begin{eqnarray}
{\cal I} &=& 
{ 1 \over \sigma_0\left(\sqrt s\, \right)}
\sum_n { 1 \over n!}\,\int d\vec p_1\cdots d\vec p_n\
\delta\left(\sum\vec p_i\right)\,
{h\!\left(\sqrt{s}\right)}\,
\nonumber\\
&&\times 
F_{\!n}\!\left(\vec p_1,\dots,\vec p_n,\sqrt s,
\alpha_s\!\left(C \sqrt S\right)\right) 
{\cal S}_n\left(\vec p_1\sqrt{S/s},\dots ,\vec p_n\sqrt{S/s}\right).
\label{integrationstruct}
\end{eqnarray}

Eq.~(\ref{integrationstruct}) is implemented as an event generator.
Events with $n$ partons with scaled momenta $\vec p_i$ are generated
along with a weight equal to $h\, F_n/\sigma_0$ in
Eq.~(\ref{integrationstruct}) divided by the density of points in $\vec
p_i$ space. A separate routine then multiplies the weights by the
measurement function ${\cal S}_n\left(\vec p_1[{S/s}]^{1/2},\dots ,\vec
p_n[{S/s}]^{1/2}\right)$ and takes the average of these results over a
large number of generated points.

Two features are especially worth noting in the main part of  
Eq.~(\ref{integrationstruct}), that is in every part other than the
measurement functions. First, the true momentum variables have been
replaced by dimensionless variables $\vec p_i$. Second, the energy
conserving delta function has been replaced by $h(\sqrt s)$, so that the
dimensionless energy $\sqrt s$ is not fixed. The true c.m. energy
$\sqrt S$ appears in only one place, in the argument of $\alpha_s$.  In
this calculation, we are using the massless theory. However, if one
wanted to add quark masses $m_i$, then the functions $F_{\!n}$ in
Eq.~(\ref{integrationstruct}) would depend on additional dimensionless
parameters $[s/S]^{1/2}\,m_i$. Then the c.m. energy $\sqrt S$
would appear in the argument of $\alpha_s$ and in these dimensionless
mass parameters.

\begin{figure}
\includegraphics[width = 8 cm]{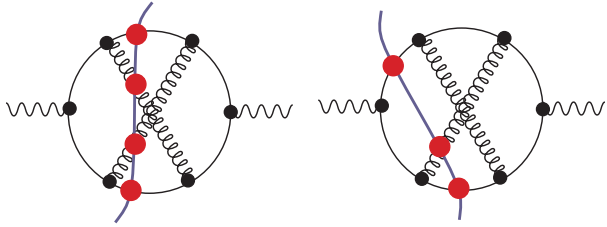}
\medskip
\caption{Two cuts of one of the Feynman diagrams that contribute to 
$e^+e^- \to {\it hadrons}$.}
\label{fig:cutdiagrams}
\end{figure}

The contribution ${\cal I}^{(2)}$ to $\cal I$ proportional to
$\alpha_s^2$ can be expressed in terms of cut Feynman diagrams, as in
Fig.~\ref{fig:cutdiagrams}. (In this section, we consider diagrams that
do not have self-energy subdiagrams, since self-energy diagrams require
a special treatment.) The dots where the parton lines cross the
cut represent the function ${\cal S}_n$. Each diagram is a three loop
diagram, so we have integrations over loop momenta $\vec l_1$,
$\vec l_2$ and $\vec l_3$. Eq.~(\ref{integrationstruct}) lacks an
energy conserving delta function, so we have integrations over four
energies, which we might take to be loop energies $l_1^0$,
$l_2^0$ and $l_3^0$ chosen in the same way as the loop momenta and the
energy  $l_0^0$ entering the graph on the vector boson line. We first
perform the energy integrations. For the graphs in which four parton
lines cross the cut, there are four mass-shell delta functions
$\delta (p_J^2)$. These delta functions eliminate the three energy
integrals over $l_1^0$, $l_2^0$, and $l_3^0$ as well as the integral
over $l_0^0$. For the graphs in which three parton lines cross the cut,
we can eliminate the integration over $l_0^0$ and two of the $l_J^0$
integrals. One integral over the energy $E$ in the virtual loop remains.
To perform this integration, we close the integration contour in the
lower half $E$ plane. This gives a sum of terms obtained from the
original integrand by some simple algebraic substitutions in which $E$
is replaced by a location $E_i$ of one of the poles in the lower half
$E$ plane. Then we do the same thing except that we close the
integration contour in the upper half $E$ plane. Finally, we take the
average of these results. (For well behaved integrands, these two
contributions are the same, but in Coulomb gauge some of the integrands
are not so well behaved, as we shall see.)

Having performed the energy integrations, we are left with an integral
of the form
\begin{equation}
{\cal I}^{(2)} = 
\sum_G
\int d \vec l_1\,d \vec l_2\,d \vec l_3\
\sum_C\,
g(G,C;\vec l_1,\vec l_2,\vec l_3).
\label{master}
\end{equation}
Here there is a sum over graphs $G$ (of which one is shown in
Fig.~\ref{fig:cutdiagrams}) and there is a sum over the possible cuts
$C$ of a given graph. The problem of calculating ${\cal I}^{(2)}$ is
now set up in a convenient form for calculation.

If we were using the Ellis-Ross-Terrano method for doing
next-to-leading order calculations \cite{ERT}, we would put the sum over
cuts outside of the integrals in Eq.~(\ref{master}). For those cuts $C$
that  have three partons in the final state, there is a virtual loop.
We can arrange that one of the loop momenta, say $\vec l_1$, goes around
this virtual loop. The essence of the Ellis-Ross-Terrano method is to
perform the integration over the virtual loop momentum analytically,
while the remaining integrations are performed numerically.
The integration over the virtual loop momentum is often ultraviolet
divergent, but the ultraviolet divergence is easily removed by a
renormalization subtraction. The integration is also typically infrared
divergent. This divergence is regulated by working in $3 - 2\epsilon$
space dimensions and then taking $\epsilon \to 0$ while dropping the
$1/\epsilon^n$ contributions (after proving that they cancel against
other contributions). After the $\vec l_1$ integration has been
performed analytically, the integrations over $\vec l_2$ and $\vec l_3$
can be performed numerically. For the cuts $C$ that have four partons
in the final state, there are also $1/\epsilon^n$ infrared divergences.
One uses either a ``phase space slicing'' or a ``subtraction''
procedure to isolate these divergences, which cancel the
$1/\epsilon^n$ pieces from the virtual graphs. In the end, we are left
with an integral $\int d\vec l_1\,d\vec l_2\,d\vec l_3$ in exactly
three space dimensions that can be performed numerically.

In the numerical method, we keep the sum over cuts $C$ inside the
integrations. We take care of the ultraviolet divergences by simple
renormalization subtractions on the integrand. We make certain
deformations on the integration contours so as to keep away from poles
of the form $1/[E_F - E_I \pm i\epsilon]$, where $E_F$ is the energy of
the final state and $E_I$ is the energy of an intermediate state. Then
the integrals are all convergent and we calculate them by Monte Carlo
numerical integration.

Let us now look at the contour deformation in a little more detail. We
denote the momenta $\{\vec l_1,\vec l_2,\vec l_3\}$ collectively
by $ l$ whenever we do not need a more detailed description. Thus
%*
\begin{equation}
{\cal I}^{(2)} = 
\sum_G
\int\! d l\
\sum_C\,
g(G,C; l).
\label{basicagain}
\end{equation}
For cuts $C$ that leave a virtual loop integration, there are
singularities in the integrand of the form $E_F - E_I +i\epsilon$ (or
$E_F - E_I -i\epsilon$ if the loop is in the complex conjugate
amplitude to the right of the cut). Here $E_F$ is the energy of the
final state defined by the cut $C$ and $E_I$ is the energy of a
possible intermediate state. For the purpose of this review, all we need
to know is that $E_F - E_I = 0$ on a surface in the space of $\vec l_1$
for fixed $\vec l_2$ and $\vec l_3$ if we pick $\vec l_1$ to be the
momentum that flows around the virtual loop. These singularities do not
create divergences. The Feynman rules provide us with the $i\epsilon$
prescriptions that tell us what to do about the singularities: we
should deform the integration contour into the complex $\vec l_1$ space
so as to keep away from them. Thus we write our integral in the form
%*
\begin{equation}
{\cal I}^{(2)} = 
\sum_G
\int\! d l\
\sum_C\,
{\cal J}(G,C; l)\,
g(G,C; l+i\kappa(G,C; l)).
\label{deformed}
\end{equation}
Here $i\kappa$ is a purely imaginary nine-dimensional vector that we add
to the real nine-dimensional vector $ l$ to make a complex
nine-dimensional vector. The imaginary part $\kappa$ depends on the
real part $ l$, so that when we integrate over $ l$, the
complex vector $ l + i\kappa$ lies on a surface, the integration
contour, that is moved away from the real subspace. When we thus deform
the contour, we supply a jacobian ${\cal J} = \det(\partial ( l +
i\kappa)/\partial  l)$. (See Ref.~\cite{beowulfPRD} for details.) 

The amount of deformation $\kappa$ depends on the graph $G$ and,
more significantly, the cut $C$. For cuts $C$ that leave no virtual
loop, each of the momenta $ \vec l_1$, $ \vec l_2$, and $\vec
 l_3$ flows through the final state. For practical reasons, we want
the final state momenta to be real. Thus we set $\kappa = 0$ for cuts
$C$ that leave no virtual loop. On the other hand, when the cut $C$
does leave a virtual loop, we choose a non-zero $\kappa$. We must,
however, be careful. When $\kappa = 0$ there are singularities in $g$
on certain surfaces that correspond to collinear parton momenta. These
singularities cancel between $g$ for one cut $C$ and $g$ for another.
This cancellation would be destroyed if, for $ l$ approaching the
collinear singularity, $\kappa = 0$ for one of these cuts but not for
the other. For this reason, we insist that for all cuts $C$, $\kappa \to
0$ as $ l$ approaches one of the collinear singularities. The details
can be found in Ref.~\cite{beowulfPRD}. All that is important here is
that $\kappa \to 0$ quadratically with the distance to a collinear
singularity.

Much has been left out in this brief overview, but we should now have
enough background to see what to do in Coulomb gauge. It might seem
that all we have to do is use the Feynman rules in Coulomb gauge, but
there are some questions associated with the two and three point
subdiagrams that need a special analysis. We now turn to that analysis,
beginning with a description of a momentum space coordinate system that
is useful for the two point subdiagrams.

\section{Elliptical coordinates}
\label{sec:elliptical}

Most of our effort will be devoted to the gluon and quark one loop
self-energy diagrams. In these diagrams, we begin by performing the
integration over the loop energy by contour integration. This leaves an
integration over the loop three-momentum. We will be greatly helped by
expressing the components of the loop momentum in terms of three
appropriately chosen variables $\{\Delta, x, \phi\}$. These
variables are defined by considering, instead of the virtual
self-energy graph, the corresponding cut self-energy graph.

Consider an off-shell parton carrying momentum $\bar q^\mu$ that
splits into two partons that carry momenta $k_\pm^\mu$. Let these two
partons be on-shell, $k_\pm^2 = 0$. Thus $k_\pm^0 = \omega_\pm	$ where
\begin{equation}
\omega_\pm \equiv |\vec k_\pm|.
\end{equation}
We consider the space-part of $\bar q^\mu$ to be fixed, and we call it
$\vec q$, while the energy is determined by energy conservation
\begin{equation}
\bar q = (\omega_+ + \omega_-, \vec q).
\end{equation}
We define a loop momentum $ \vec l$ by
\begin{equation}
\vec k_\pm = {\scriptstyle {1\over 2}}\,\vec q \pm  \vec l.
\end{equation}

We will use elliptical coordinates $\{\Delta, x, \phi\}$ defined as
follows. First, define ${\cal Q}$ by
\begin{equation}
{\cal Q} = |\vec q\,|.
\end{equation}
Now, define coordinates $\{\Delta,x\}$ by
\begin{eqnarray}
\Delta + 1 &=& { 1 \over {\cal Q}}\,
\left(
\omega_+ + \omega_-
\right),
\nonumber\\
2 x - 1  &=& { 1 \over {\cal Q}}\,
\left(
\omega_+ - \omega_-
\right).
\end{eqnarray}
Then
\begin{equation}
0 < \Delta 
,\hskip 1 cm 
0 < x < 1.
\end{equation}
Finally, let $\phi$ be the azimuthal angle of $\vec l$ in a
coordinate system in which the $z$-axis lies in the direction of $\vec
q$ and the direction of the $x$-axis is defined arbitrarily.
Often, we will want to work in $3 - 2\epsilon$ space dimensions. In
this case, $\phi$ stands for a point on the surface of a unit sphere in
$2 - 2
\epsilon$ dimensions, with
\begin{equation}
\int d\phi \equiv S(2 - 2\epsilon)
= { 2 \pi^{1-\epsilon} \over \Gamma(1-\epsilon)}.
\end{equation}

The surfaces of constant $\Delta$ are ellipsoids, while the surfaces of
constant $x$ are paraboloids that are orthogonal to the constant
$\Delta$ surfaces. Both of these surfaces are orthogonal to the
surfaces of constant $\phi$.

We note immediately that 
\begin{equation}
\bar q^2 = (\omega_+ + \omega_-)^2 - {\cal Q}^2
= {\cal Q}^2
\left[ (\Delta + 1)^2 - 1 
\right]
= {\cal Q}^2 \Delta (\Delta + 2).
\end{equation}
The inverse relation is
\begin{equation}
\Delta = \sqrt{1 + \bar q^2/{\cal Q}^2} - 1.
\end{equation}
It will often be convenient to use $\bar q^2$ as an independent variable
instead of $\Delta$.

The part, $\vec l_T$, of $ \vec l$ transverse to $\vec q$ is determined
by a unit vector in $2 - 2 \epsilon$ dimensions specified by $\phi$
and by the magnitude $|\vec l_T|$, which is
\begin{equation}
|\vec l_T| =
\sqrt{x(1-x)\,\bar q^2}.
\end{equation}
The component of $ \vec l$ along $\vec q$ is
\begin{equation}
 \vec l \cdot \vec q /{\cal Q}
= ({\cal Q}/2)\, (1 + \Delta ) (2x-1).
\end{equation}

A straightforward calculation shows that the jacobian of the
transformation is given by
\begin{equation}
{ d^{3-2\epsilon}\vec l \over 2\omega_+2\omega_-}
=  { 1 \over 8 {\cal Q}}\,
{ 1 \over 1 + \Delta}\,
[x(1-x)\,\bar q^2]^{-\epsilon}
d \bar q^2\, dx\,  d^{1-2\epsilon}\phi.
\label{jacobian}
\end{equation}

With $\epsilon = 0$, the transformation from $\{\Delta, x,\phi\}$ to
$\vec l$ is
\begin{equation}
\vec l =
  |\vec l_T|\,\cos\phi\ \vec n_x
+ |\vec l_T|\,\sin\phi\ \vec n_y
+ ( \vec l \cdot \vec q /{\cal Q})\, \vec n_z .
\label{elltransform}
\end{equation}
Here $\vec n_z = \vec q /{\cal Q}$, while $\vec n_x$ and $\vec n_y$ are
two unit vectors orthogonal to each other in the plane orthogonal to
$n_z$.

\section{Structure of graphs with a cut gluon propagator}
\label{sec:cutgluonstructure}

In this and the following sections, we analyze the gluon
propagator. We denote by $n^\mu$ a unit vector in the time direction,
$n = (1,0,0,0)$. We use Coulomb gauge. Information about the use of
Coulomb gauge can be found in Refs.~\cite{ChristLee} and
\cite{leibrandt}. We consider diagrams in which
there is zero or one loop in the gluon propagator. Thus we deal
with the gluon propagator at orders $\alpha_s^0$ and $\alpha_s^1$.

\begin{figure}[htb]
\includegraphics[width = 6 cm]{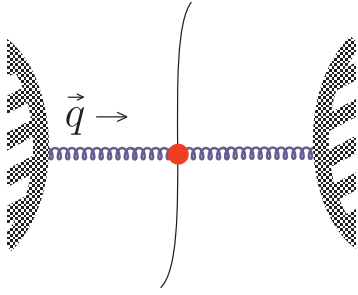}
\caption{Cut gluon propagator at the Born level.}
\label{fig:bornselfenergy}
\end{figure}

We will be interested in the factors in the cross section that arise
from a cut gluon propagator when the cut propagator is a subgraph of a
larger graph. To set the notation, we write the contribution from an
order $\alpha_s^0$ cut propagator, illustrated in
Fig.~\ref{fig:bornselfenergy}, as
\begin{equation}
{\cal I}[{\rm Born}] = \int\!d\vec q\ 
{ D(q)^{\mu\nu} \over 2{\cal Q}}\, R_{\mu\nu}^{(0)},
\label{gluonborn}
\end{equation}
where $\vec q$ is the three-momentum carried by the propagator, ${\cal
Q}\equiv |\vec q\,|$, and $q = ({\cal Q},\vec q\,)$. Then $d\vec
q/[2{\cal Q}]$ is the standard Lorentz invariant integration over the
gluon mass shell. The tensor $R_{\mu\nu}^{(0)}$ denotes the factors
associated with the rest of the graph and with the final state
measurement function ${\cal S}$; $R_{\mu\nu}^{(0)}$ depends on $\vec
q$, but this dependence is suppressed in the notation.
Finally, $D^{\mu\nu}$ is the numerator
function for a bare gluon propagator in Coulomb gauge, which is (for
both on-shell and off-shell gluons),
\begin{equation}
D(k)^{\mu\nu} =
-g^{\mu\nu}
+{ 1 \over \omega^2}
\left[
-k^\mu \tilde k^\nu
-\tilde k^\mu k^\nu
+ k^\mu k^\nu
\right],
\label{Dmunu}
\end{equation}
where
\begin{equation}
\tilde k = (0,\vec k)
\end{equation}
consists of just the $\mu = 1,2$ and 3 components of $k^\mu$ and where,
as in the previous section,
\begin{equation}
\omega = |\vec k| = \sqrt{-\tilde k ^2}.
\end{equation}

\begin{figure}[htb]
\includegraphics[width = 6 cm]{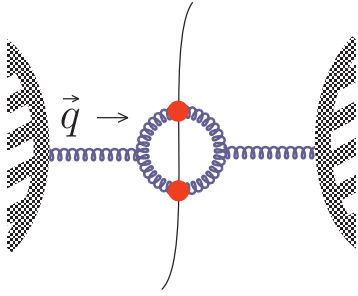}
\caption{Cut gluon self-energy diagram.}
\label{fig:cutselfenergy}
\end{figure}

At order $\alpha_s^1$, we consider a gluon propagator with a one
loop self-energy subgraph. We consider three cases. In the first case,
the two bare propagators in the self-energy subgraph are cut and the
neighboring bare propagators are virtual. This case is illustrated in
Fig.~\ref{fig:cutselfenergy}. We write the contribution of a cut gluon
self-energy graph in the form
\begin{equation}
{\cal I}[{\rm real}] = \int\!d\vec q\
{ 1 \over 4 \pi {\cal Q}}\int 
{d\bar q^2\over \bar q^2}\,dx\,d\phi\ 
{\cal M}_g^{\mu\nu}(\bar q^2,x,\phi)\,
R_{\mu\nu}(\bar q^2,x,\phi),
\label{Ireal}
\end{equation}
ignoring the infrared divergence, which will cancel inside the integral
after combining real and virtual contributions in
Eq.~(\ref{Igrealandvirt}) below. The integration over the loop momentum
$ \vec l$ has been changed to an integration over $\{\bar q^2,x,\phi\}$
and the factors coming from the cut self-energy graph and its adjacent
virtual gluon propagators are included in ${\cal M}/(4\pi{\cal Q}\bar
q^2)$. Then $R_{\mu\nu}$ represents the rest of the graph, including
the measurement function ${\cal S}$. All of the factors in
$R_{\mu\nu}$ depend on the virtuality $\bar q^2$. The
measurement function depends also on $x$ and $\phi$. The dependence of
$R_{\mu\nu}$ on $\vec q$ is suppressed in the notation. The
conventions chosen are such that in Eq.~(\ref{gluonborn}) for a cut
bare propagator, $R_{\mu\nu}^{(0)}$ is  $R_{\mu\nu}(\bar q^2,x,\phi)$
evaluated with $\bar q^2 = 0$. Here we use the infrared safety
property (\ref{IRsafety}) of the measurement function.

\begin{figure}[htb]
\includegraphics[width = 6 cm]{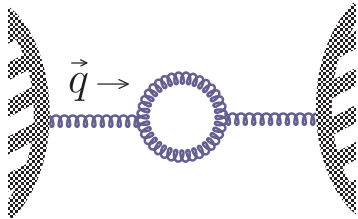}
\caption{Off-shell virtual gluon self-energy diagram. (This case does
not occur in order $\alpha_s^2$ graphs for $e^+ e^- \to hadrons$, but
we consider it as an intermediate step toward analyzing the on-shell
gluon self-energy diagram. The analogous off-shell virtual quark
self-energy diagram does occur in order $\alpha_s^2$ graphs for $e^+ e^-
\to hadrons$.)}
\label{fig:allvirtualselfenergy}
\end{figure}

In the second case to be considered below the self-energy loop is
entirely virtual and the neighboring bare propagators are not cut, so
that the incoming momentum $q$ is off-shell. This case is illustrated
in Fig.~\ref{fig:allvirtualselfenergy}. We consider $q^2 < 0$. In this
case, we investigate the quantity
\begin{equation}
F_g^{\mu\nu}(q)
={ 1 \over q^2}\
D(q)_\alpha^{\mu}\,\Pi(q)^{\alpha\beta}\,D(q)_\beta^{\nu},
\label{Fmunu}
\end{equation}
where $\Pi^{\alpha\beta}$ is the self-energy function,
Eq.~(\ref{pimunu}). There are two factors of $1/q^2$ in the
propagator function; we include just one of them in the definition of
$F_g$. We find that $F_g^{\mu\nu}(q)$ is given by an integral of the
form
\begin{equation}
F_g^{\mu\nu}(q) = { 1 \over 2 \pi}\int 
d\bar q^2\,dx\,d\phi\ 
{ 1 \over \bar q^2 - q^2}\,
{W}_g^{\mu\nu}(\bar q^2,x,\phi).
\label{Wdef}
\end{equation}
The \MSbar\ ultraviolet renormalization is expressed through certain
subtraction terms included in $W$.

\begin{figure}[htb]
\includegraphics[width = 6 cm]{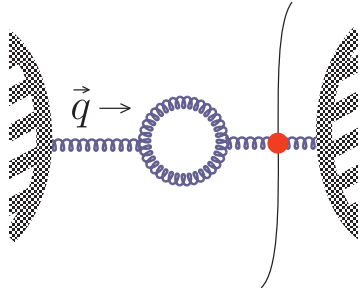}
\caption{On-shell virtual gluon self-energy diagram.}
\label{fig:virtualselfenergy}
\end{figure}

In the third case, the self-energy loop is entirely virtual and one of
the neighboring bare propagators {\em is} cut, so that the incoming
momentum  satisfies $q^2 = 0$. This case is illustrated in
Fig.~\ref{fig:virtualselfenergy}. Then we need $F_g^{\mu\nu}(q)$
with $q^2 = 0$, which takes the form
\begin{equation}
F_g^{\mu\nu}(q) = { 1 \over 2 \pi}\int 
{ d\bar q^2 \over \bar q^2}\,dx\,d\phi\ 
{\cal W}_g^{\mu\nu}(\bar q^2,x,\phi),
\label{calWdef}
\end{equation}
where ${\cal W}_{\!g}$ is $W_{\!g}$ with $q^2 = 0$. Again, the integral
is infrared divergent, but the divergence will cancel inside the
integrand in Eq.~(\ref{Igrealandvirt}). The corresponding
contribution to the cross section ${\cal I}$ is
\begin{eqnarray}
{\cal I}[{\rm virtual}]&=& \int\!d\vec q\
{ 1 \over 2 {\cal Q}}\,
F_g^{\mu\nu}(q)\,
R_{\mu\nu}^{(0)}
\label{Ivirtg}
\\
&=&\int\!d\vec q\
{ 1 \over 4 \pi {\cal Q}}\int {d\bar q^2\over \bar q^2}\,dx\,d\phi\
{\cal W}_g^{\mu\nu}(\bar q^2,x,\phi)\,
R_{\mu\nu}^{(0)}.
\nonumber
\end{eqnarray}
We should note that ${\cal I}[{\rm virtual}]$ in Eq.~(\ref{Ivirtg}) is
the total contribution from the two graphs in which one or the other of
the neighboring bare propagators is cut. The contribution from either
of these graphs is half of this.

When we add the contributions from graphs with a cut self-energy
subdiagram and with a virtual self-energy subdiagram with a cut
adjoining bare propagator, we get
\begin{eqnarray}
{\cal I}[{\rm real}]+{\cal I}[{\rm virtual}]&=&
\int\!d\vec q\
{ 1 \over 4 \pi {\cal Q}}\int {d\bar q^2\over \bar q^2}\,dx\,d\phi\
\label{Igrealandvirt}\\
&&\times\left\{
{\cal M}_g^{\mu\nu}(\bar q^2,x,\phi)\,
R_{\mu\nu}(\bar q^2,x,\phi)
+
{\cal W}_g^{\mu\nu}(\bar q^2,x,\phi)\,
R_{\mu\nu}^{(0)}
\right\}.
\nonumber
\end{eqnarray}
As noted above,
the integrals of the two terms separately would be infrared divergent
and would not make sense by themselves. However, the $\bar q^2 \to 0$
singularities in the integrand in Eq.~(\ref{Igrealandvirt}) cancel, so
that the integral is finite and is suitable for calculation by
numerical integration.

With this preparation, we are ready to turn to the calculations for the
three cases.

\section{Real gluon self-energy graph}
\label{sec:realglue}

It is straightforward to evaluate the function ${\cal M}_g^{\mu\nu}$
defined in Eq.~(\ref{Ireal}). We find
\begin{eqnarray}
{\cal M}_g^{\mu\nu}&=& { \alpha_s \over 4\pi}\,{ 1 \over 1 + \Delta}
\nonumber\\
&&\times\Biggl\{N_{TT}\,D({\cal Q},\vec q)^{\mu\nu}
+N_{tt}\left[
{  l_T^\mu l_T^\nu \over \vec l_T^{\,2}}
- {\textstyle{1 \over 2}}D({\cal Q},\vec q)^{\mu\nu}
\right]
\nonumber\\
&&
+{ \bar q^2 \over {\cal Q}^2}N_{EE}\,n^\mu n^\nu
+ N_{Et}{ \bar q\cdot n \over (1+\Delta){\cal Q}^2}
\left( l_T^\mu n^\nu + n^\mu  l_T^\nu \right)
\Biggr\}.
\label{calMgterms}
\end{eqnarray}
The coefficients $N_{TT},N_{tt}, N_{EE},$ and $N_{Et}$ are functions of
$\bar q^2$ and $x$ and are given below. The tensor $D({\cal Q},\vec
q)^{\mu\nu}$ is the numerator (\ref{Dmunu}) for an on-shell gluon,
with momentum $q_{\rm os} = ({\cal Q},\vec q)$. In a reference frame
in which $\vec q$ is aligned with the $z$-axis, the only non-zero
components of $D({\cal Q},\vec q)^{\mu\nu}$ are  $D({\cal Q},\vec
q)^{ij} = \delta^{ij}$ with $i,j\in \{1,2\}$. In the second term, using
the notation of Sec.~\ref{sec:elliptical}, $ l_T^\mu$ is the vector
\begin{eqnarray}
 l_T^0 &=& 0,
\nonumber\\
 \vec l_T &=&  \vec l 
- {  \vec l \cdot \vec q \over {\cal Q}^2}\ \vec q,
\end{eqnarray}
or
\begin{equation}
 l_T^\mu = \sqrt{x(1-x)\,\bar q^2 }
\left(\cos\phi\,n_x^\mu + \sin\phi\,n_y^\mu\right).
\end{equation}
Note that this term vanishes if we average over angles $\phi$. The
third term gives ${\cal M}^{00}$. It vanishes when $\bar q^2
\to 0$. Finally, the fourth term gives ${\cal M}^{0i}$ and ${\cal
M}^{i0}$. Note that it vanishes if we average over angles $\phi$.

We have written Eq.~(\ref{calMgterms}) in a more elaborate form than
might have seemed necessary: $\bar q \cdot n = (1+\Delta)Q$, so the
coefficient in the fourth term could have been simplified. The reason
for the more elaborate form is as follows. The tensor ${\cal
M}_g^{\mu\nu}$ is a function of the momenta $\bar q^\mu, k_+^\mu,
k_-^\mu$ carried on the lines of the graph. In the derivation, we have
understood that gluons with momenta $k_+^\mu$ and $k_-^\mu$ enter the
final state. However in a calculation, it might be convenient to use
momenta with reversed signs: $\bar q^\mu \to -\bar q^\mu, k_+^\mu \to -
k_+^\mu, k_-^\mu \to -k_-^\mu$. Then the vector $ l_T^\mu$ is also
reversed, while $\cal Q$, $\Delta$, and $x$ are unchanged. We have
written ${\cal M}_g^{\mu\nu}$ in Eq.~(\ref{calMgterms}) in a form that
is unchanged under this reversal of momenta.

The coefficients are 
\begin{eqnarray}
N_{TT} &=&
2 C_A\biggl\{
x(1-x) + { 8 x(1-x)[1- x(1-x)] \over 
\bar q^2/{\cal Q}^2 + 4x(1-x)}
\nonumber\\
&&\hskip 1 cm + { 16 x(1-x)[1- 4x(1-x) + 2x^2(1-x)^2] \over 
[\bar q^2/{\cal Q}^2 + 4x(1-x)]^2}
\biggr\}
\nonumber\\
&&+ N_F\biggl\{
1 - 2 x(1-x)
\biggr\},
\nonumber\\
N_{tt} &=&
4 C_A\biggl\{
x(1-x) - { 8 x^2(1-x)^2 \over 
\bar q^2/{\cal Q}^2 + 4x(1-x)}
 + { 32 x^3(1-x)^3 \over 
[\bar q^2/{\cal Q}^2 + 4x(1-x)]^2}
\biggr\}
\nonumber\\
&&- 4 N_F x(1-x),
\nonumber\\
N_{EE} &=&
C_A\biggl\{
[1-4x(1-x)] - { 8 x(1-x)[1 - 4x(1-x)] \over 
\bar q^2/{\cal Q}^2 + 4x(1-x)}
\nonumber\\
&&\hskip 1 cm + { 32 x^2(1-x)^2[1- 4x(1-x)] \over 
[\bar q^2/{\cal Q}^2 + 4x(1-x)]^2}
\biggr\}
\nonumber\\
&&+ 4 N_F x(1-x),
\nonumber\\
N_{Et} &=&
2 C_A (2x-1)\biggl\{
-1 - { 2[1 - 4x(1-x)] \over 
\bar q^2/{\cal Q}^2 + 4x(1-x)} 
+ {16 x(1-x)[1- 2x(1-x)] \over 
[\bar q^2/{\cal Q}^2 + 4x(1-x)]^2}
\biggr\}
\nonumber\\
&& +2 N_F (2x-1).
\label{calMcoefs}
\end{eqnarray}
(In this paper, we use the standard notation in which $N_F$ is the
number of quark flavors, $N_C$ is the number of colors, $C_A = N_C$,
and $C_F = (N_C^2 - 1)/(2N_C)$.)

The behavior of ${\cal M}^{\mu\nu}$ as $\bar q^2 \to 0$ is important.
Leaving out the $N_{tt}$ and $N_{Et}$ terms, which average to zero
after integrating over angles, we obtain
\begin{equation}
{\cal M^{\mu\nu}} \sim 
D(q)^{\mu\nu}\,{ \alpha_s \over 2\pi}
\left\{
{\textstyle{1\over 2}}\,\tilde P_{g/g}(x)
+ N_F \tilde P_{q/g}(x)
\right\},
\label{calM0}
\end{equation}
where
\begin{eqnarray}
\tilde P_{q/g}(x)&=&{\textstyle{1\over 2}}[1-2 x(1-x)],
\nonumber\\
\tilde P_{g/g}(x)&=& 2C_A
{ [1- x(1-x)]^2 \over x(1-x)}
\label{APkernels}
\end{eqnarray}
are the one loop parton evolution kernels without their $x \to 1$
regulation. Notice that $\tilde P_{g/g}(x)$ is singular at $x \to 0$
and $x \to 1$ but that ${\cal M}^{\mu\nu}$ is {\em not} singular at
these points. The singularities emerge only when we take the $\bar q^2
\to 0$ limit. Also notice that in front of $\tilde P_{g/g}(x)$ in
Eq.~(\ref{calM0}) there is a symmetry factor $1/2$, which arises
because the two final state gluons in $g \to g g$ are identical.

\section{Virtual gluon self-energy graph}
\label{sec:virtualglue}

In this section, we analyze the virtual gluon self-energy graph at
spacelike momentum $q^\mu$. We consider the self-energy
function $\Pi(q)^{\mu\nu}$. Later, we also consider the quantity 
\begin{equation}
F_g^{\mu\nu}(q)
={ 1 \over q^2}\
D(q)_{\mu'}^{\mu}\,\Pi(q)^{\mu'\nu'}\,D(q)_{\nu'}^{\nu},
\label{Fgmunu}
\end{equation}
where $D(q)_{\mu'}^{\mu}$ is the numerator (\ref{Dmunu}) of the bare
gluon propagator. In applications with a cut gluon propagator, as
in Fig.~\ref{fig:virtualselfenergy}, one needs
$F_g^{\mu\nu}(q)$ evaluated with $q^2 = 0$.

We let the momenta of the two partons in the loop be $k_\pm^\mu$, with
$k_+^\mu + k_-^\mu = q^\mu$. Then the Feynman rules for
$\Pi(q)^{\mu\nu}$ give
\begin{eqnarray}
\Pi(q)^{\mu\nu}&=&
ig^2\,\tilde\mu^{2\epsilon}
\int{ d^{4-2\epsilon}k_+\over (2\pi)^{4-2\epsilon}}\
{ 1 \over (k_+^2 + i\epsilon)(k_-^2+i\epsilon)}
\nonumber\\
&&\times\biggl\{
- {C_A\over 2}\,
V^{\mu\alpha\beta}(q,-k_+,-k_-)\,
D_{\alpha\alpha'}(k_+)\,
D_{\beta\beta'}(k_-)\,
V^{\alpha'\beta'\nu}(k_+,k_-,-q)
\nonumber\\
&&\ \ \ \ +{C_A\over 2}\,
(\tilde k_+^\mu \tilde k_-^\nu + \tilde k_-^\mu \tilde k_+^\nu)\
{ k_+^2 k_-^2 \over \omega_+^2 \omega_-^2}
\nonumber\\
&&\ \ \ \ + {1 \over 2}\,N_F\,{\rm Tr}\left[
\gamma^\mu \rlap{/}k_-
\gamma^\nu \rlap{/}k_+
\right]
\biggr\}
\nonumber\\
&& 
- ({\rm pole}),
\label{pimunu}
\end{eqnarray}
where
\begin{equation}
V^{\alpha\beta\gamma}(k_A,k_B,k_C) =
 g^{\alpha\beta} (k_A^\gamma -k_B^\gamma)
+g^{\beta\gamma} (k_B^\alpha -k_C^\alpha)
+g^{\gamma\alpha}(k_C^\beta  -k_A^\beta ).
\end{equation}
Here the first line is for the gluon loop, the second line is for the
ghost loop, symmetrized over the two possible momentum labellings, and
the third line is for the quark loop. The gluon loop includes a
symmetry factor $1/2$. The terms corresponding to gluon and ghost loops
contain group factors $C_A = 3$, while the term corresponding to a
quark loop has a factor $N_F$ and a color factor $T_F = 1/2$. We
subtract the ultraviolet pole, as required by the \MSbar\ prescription.
(The parameter $\tilde \mu$ is related to the \MSbar\ scale $\mu$ by
$\mu^2 = 4\pi\tilde\mu^2 e^{-\gamma}$.)

We choose a coordinate system in which the $z$-axis is aligned with
$\vec q$.  Then $q = (q^0,0,0,{\cal Q})$. We assume that $q^2 <0$. Later
we will take the limit $q^2 \to 0$.

\subsection{The energy integral}
\label{sec:energyintegrals}

We begin by performing the $k_+^0$ integral. Consider a term in
$\Pi^{\mu\nu}$ of the form
\begin{equation}
ig^2\,\tilde\mu^{2\epsilon}
\int{ d^{4-2\epsilon}k_+\over (2\pi)^{4-2\epsilon}}\
{A\, f(k_+,k_-) 
\over (k_+^2 + i\epsilon)(k_-^2+i\epsilon)},
\label{fdefinition}
\end{equation}
where $A$ is a function of $\vec k_+$ and $\vec k_-$ but is
independent of $k_+^0$ and $k_-^0$. We perform the
$k_+^0$ integration, leaving an integral over $\vec k_+$. Then we
change variables to the elliptical coordinates $\{\bar q^2, x, \phi\}$ 
defined from $\vec k_\pm$ as in Sec.~\ref{sec:elliptical}, so that our
term takes the form
\begin{equation}
-{ \alpha_s \over 8\pi^2}
\int d\bar q^2\,dx\,d^{1-2\epsilon}\!\phi\
\left[x(1-x)\bar q^2 \over 4\pi^2\tilde\mu^2\right]^{-\epsilon}
{A\, g(\bar q^2,x,\phi) 
\over \bar q^2 - q^2}.
\label{gdefinition}
\end{equation}
Thus we will need an integration table that translates $f$ into $g$.

There is, however, a problem. Some of the integrals over $k_+^0$ are
divergent. Thus we need a definition. We elect to perform the
integrals over loop energies inside the integrals over loop
three-momenta. We calculate these integrals by closing the energy
contours in the lower half plane and then calculate them again by
closing the contours in the upper half plane. Finally, we average the
two results. The radius of the large semicircles that close the
contours are always to be big enough to enclose all poles. Thus our
prescription is a simple algebraic prescription of adding, with the
appropriate signs and a factor 1/2, the residues of all the poles in
the complex energy plane. If we apply this prescription in Feynman
gauge, then the energy integrations are all convergent and we get the
usual answer. In Coulomb gauge, we make this prescription part of the
definition of the gauge.  The required integral table is given as Table
\ref{tab:integraltable}.

\begin{table}[htb]
\caption{\label{tab:integraltable}
Integral table relating $f$ in Eq.~(\ref{fdefinition}) 
to $g$ in Eq.~(\ref{gdefinition}).}
\begin{ruledtabular}
\begin{tabular}{cc}
$f$&$g$\\
\colrule
1&1\\[10pt]
$k_+^2 - k_-^2$
&$(q^2 - \bar q^2)\,
\displaystyle{ 2 x - 1 \over 1+\Delta}$\\[10pt]
$k_+^2 + k_-^2$
&$(q^2 - \bar q^2)$\\[10pt]
$(k_+^2 - k_-^2)^2$
&$(q^2 - \bar q^2)\,(q^2 + 4 {\cal Q}^2 x(1-x))$\\[10pt]
$(k_+^2)^2 - (k_-^2)^2$
&$(q^2 - \bar q^2)^2\
\displaystyle{ 2x-1 \over 1+\Delta}$\\[10pt]
$k_+^2 k_-^2$
&$0$\\
\end{tabular}
\end{ruledtabular}
\end{table}

Of these integrals, the most divergent is that for $f = k_+^2 k_-^2$.
There are no poles in the complex loop-energy plane, so we have
\begin{equation}
\int dE = 0.
\label{bad1}
\end{equation}
One might expect that if the  self-energy diagram is embedded in a
larger diagram as part of a gauge invariant calculation, then an 
$\int\!dE$ contribution from some other virtual loop diagram would
cancel this contribution. However, so far as we can see, this does not
happen. Specifically, the gluon-quark-antiquark one loop three point
function does not have an $\int\!dE$ divergence. Instead, we look to
the coefficient of $\int\!dE$. This coefficient is independent of $q^2$
and is proportional to the integral
\begin{equation}
\tilde\mu^{2\epsilon}\int 
{ d^{3-2\epsilon}\vec k_+ \over (2\pi)^{3-2\epsilon}}
\left\{ {1 \over \vec k_+^2} +{1 \over \vec k_-^2}\right\}.
\label{badguy}
\end{equation}
This integral vanishes in dimensional regulation for any $\epsilon$.
Thus we have an ambiguous contribution of infinity times zero and we
find it sensible that the contour integration prescription given
above instructs us to discard this contribution.

The integral for $f = (k_+^2)^2 - (k_-^2)^2$ is logarithmically
divergent and can be obtained from the simple integral
\begin{equation}
\int dE\ { E \over E^2 - M^2 +i\epsilon} = 0.
\label{bad2}
\end{equation}
Zero is a sensible result for this integral since the integrand is odd
under $E \to - E$. In our prescription, we get $-i\pi$ when we close
the contour in the lower half plane and $+i\pi$ when we close the
contour in the upper half plane. Averaging these two results give zero.

Our prescription may be compared to that of Leibbrandt and Williams
\cite{leibrandt}. In that prescription, one imagines that the loop
energy is integrated over the imaginary axis and that the integration is
performed in $1-2\beta$ dimensions. This prescription gives the same
result as the one adopted here: the integral in Eq.~(\ref{bad1})
becomes $\int d^{1-2\beta}\vec k$, which vanishes because it has no
scale, while the integral in Eq.~(\ref{bad2}) becomes $\int
d^{1-2\beta}\vec k\ \vec k\cdot \vec n/(\vec k^2 + M^2 + i\epsilon)$,
which vanishes because it is odd under $\vec k\cdot \vec n \to - \vec
k\cdot \vec n$.

\subsection{Components needed}

We will calculate the individual components of $\Pi(q)^{\mu\nu}$ that
we need in the coordinate frame with $q = (q^0,0,0,{\cal Q})$.
Since $D(q)^{\mu}_3 = D(q)^{\nu}_3 = 0$, we need not consider the
components $\Pi(q)^{3\nu}$ or $\Pi(q)^{\mu 3}$. In addition, 
$\Pi(q)^{0i} = \Pi(q)^{i0} = 0$ for $i \in \{1,2\}$ because of
rotational symmetry. Thus what we need are $\Pi(q)^{00}$ and
$\Pi(q)^{ij}$ for $i,j \in \{1,2\}$. We define
\begin{eqnarray}
\Pi(q)^{ij}&=& q^2 A_T(q^2)\, \delta^{ij} ,
\hskip 1 cm i,j \in \{1,2\},
\nonumber\\
\Pi(q)^{00}&=&{\cal Q}^2 A_E(q^2).
\end{eqnarray}
We turn first to $A_T(q)$.

\subsection{Transverse components}

We evaluate $A_T(q^2)$ using the integrals in Table
\ref{tab:integraltable} and find
\begin{eqnarray}
A_T(q^2) &=&
-{ \alpha_s \over 8\pi^2}
\int d\bar q^2\,dx\,d^{1-2\epsilon}\!\phi\
\left[x(1-x)\bar q^2 \over 4\pi^2\tilde\mu^2\right]^{-\epsilon}
{1\over \bar q^2 - q^2}
\sum_{J=0}^2\
{ A^{\prime\prime}_{T,J} \over [\bar q^2/{\cal Q}^2 + 4 x(1-x)]^J}
\nonumber\\
&& - ({\rm pole}),
\label{ATform}
\end{eqnarray}
where
\begin{eqnarray}
A^{\prime\prime}_{T,0}&=&
C_A\,{ 1 \over q^2}\biggl\{
- \bar q^2/2
- 3 q^2/2 
- 4\,{\cal  Q}^2
+ 2 \bar q^2\, x(1-x)
+ 24\,{\cal  Q}^2\,x(1-x) 
\nonumber\\
&&+{x(1-x)\over (1-\epsilon)}
\Bigl[ 
\bar q^2 
+ 4\,{\cal  Q}^2
+ 3\,q^2 
- 20\,{\cal  Q}^2\,x(1-x)
\Bigr]\biggr\}
\nonumber\\
&&
-N_F\,{ 1 \over q^2}\biggl\{
-\bar q^2 
+{x(1-x)\over (1-\epsilon)}\ 2\bar q^2
\biggr\},
\nonumber\\
A^{\prime\prime}_{T,1}&=&
C_A\,{ 1 \over q^2}\biggl\{
4\,q^2 
+ 2\,({q^2})^2/{\cal Q}^2 
+ 48\,{\cal Q}^2\,x(1-x) 
- 224\,{\cal Q}^2\,[x(1-x)]^2 
\nonumber\\
&& 
+\frac{x(1-x)}
{(1-\epsilon)}
\Bigl[ 
- 4\,q^2
- 4\,(q^2)^2 /{\cal Q}^2
- 32\,{\cal Q}^2\,{x(1-x)} 
\nonumber\\
&&\ \ \
- 28\,q^2\,{x(1-x)} 
+ 144\,{\cal Q}^2\,[x(1-x)]^2
\Bigr] \biggr\},
\nonumber\\
A^{\prime\prime}_{T,2}&=&
C_A\,{ 1 \over q^2}\biggl\{
-128\,{\cal Q}^2\,[x(1-x)]^2 
+ 512\,{\cal Q}^2\,[x(1-x)]^3 
\nonumber\\
&&
+\frac{[x(1-x)]^2}{(1-\epsilon)}
\Bigl[
  16\,q^2 
+ 16\,(q^2)^2/{\cal Q}^2
+ 64\,{\cal Q}^2\,{x(1-x)} 
\nonumber\\
&&
+ 64\,q^2\,{x(1-x)}
- 256\,{\cal Q}^2\,[x(1-x)]^2
\Bigr] \biggr\}.
\label{ATcoefs0}
\end{eqnarray}

As indicated by the notation, we expect that $q^2 A_T(q^2)$ vanishes at
$q^2 = 0$, but it is not evident from the form above that it does so.
Nevertheless, explicit analytical integration shows that $q^2 A_T(q^2)
= 0$ at $q^2 = 0$ for any $\epsilon$. In order to get a form in which
the vanishing of $q^2 A_T(q^2)$ at $q^2 = 0$ is manifest, we simply
subtract $q^2 A_T(q^2)$ at $q^2 = 0$, which is zero, from $q^2
A_T(q^2)$. Then $A_T(q^2)$ still has the form (\ref{ATform}) but
with new coefficients $A^{\prime}_{T,J}$:
\begin{eqnarray}
A^{\prime}_{T,0}&=&
2\,C_A \biggl\{ -1 + x(1-x)
+{2 x(1-x)\over (1-\epsilon)}
\biggr\}
+ N_F 
\biggl\{
1
-{2 x(1-x)\over (1-\epsilon)}
 \biggr\},
\nonumber\\
A^{\prime}_{T,1}&=&
2\,C_A \biggl\{ q^2/{\cal Q}^2 + 12 x(1-x)
-\frac{2x(1-x)}{(1-\epsilon)}
\Bigl[ 
q^2/{\cal Q}^2 + 12 x(1-x)
\Bigr]
\biggr\} ,
\nonumber\\
A^{\prime}_{T,2}&=&
16\,C_A\,x(1-x) \biggl\{2  - 8 x(1-x)
+\frac{x(1-x)}{(1-\epsilon)}
\Bigl[
q^2/{\cal Q}^2 + 8 x(1-x)
\Bigr]
\biggr\} .
\label{ATcoefs1}
\end{eqnarray}

We now examine the ultraviolet renormalization of $A_T(q^2)$. Define
\begin{eqnarray}
\Delta A_T(q^2) &=&
{ \alpha_s \over 8 \pi^2}
\int d\bar q^2\, dx\, d^{1-2\epsilon}\phi
\left[
x(1-x) \bar q^2 \over 4 \pi^2 \tilde \mu^2
\right]^{-\epsilon}
\nonumber\\
&& \times\left\{
{ -2C_A + N_F
 \over \bar q^2 + e^{\lambda_1}\mu^2}
+
x(1-x){ 2 C_A \over \bar q^2 + e^{\lambda_2}\mu^2}
+
{ x(1-x) \over 1-\epsilon}\,
{ 4 C_A - 2 N_F
 \over \bar q^2 + e^{\lambda_3}\mu^2}
\right\}
\nonumber\\
&&-
({\rm pole}).
\end{eqnarray}
Here $\mu^2 = 4\pi\tilde\mu^2\,e^{-\gamma}$ is the \MSbar\ scale,
$\gamma = 0.577\dots$ is the Euler constant, and 
$\lambda_1$, $\lambda_2$, and $\lambda_3$ are parameters that we will
adjust. The integrand in $\Delta A_T$ matches that of $A_T$ when $\bar
q^2 \to \infty$ but it has the opposite sign, so if we add $\Delta A_T$
to $A_T$, we will obtain an ultraviolet convergent integral.
Furthermore, since the integrands match for $\bar q^2 \to \infty$,
the ultraviolet pole terms that are included in the definitions are
opposite and will cancel in the sum $A_T + \Delta A_T$. We can easily
perform the integral:
\begin{eqnarray}
\Delta A_T(q^2) &=&
{ \alpha_s \over 4 \pi}\,
\Gamma(\epsilon)\,
e^{\gamma\epsilon}
\biggl\{
 -(2 C_A - N_F)
{ \Gamma(1-\epsilon)^2 \over \Gamma(2 - 2\epsilon)}\
e^{-\lambda_1\epsilon}
+ 2 C_A
{ \Gamma(2-\epsilon)^2 \over \Gamma(4 - 2\epsilon)}
e^{-\lambda_2\epsilon}
\nonumber\\
&& 
+
(4 C_A - 2 N_F)\,
{ \Gamma(2-\epsilon)^2 \over (1-\epsilon)\Gamma(4 - 2\epsilon)}
e^{-\lambda_3\epsilon}
\biggr\}
- ({\rm pole})
\nonumber\\
&=&
{ \alpha_s \over 4 \pi}
\left\{
 - (2C_A - N_F)\,
\left(2-\lambda_1\right)
+
{ 2 C_A \over 6}\,\left({ 5 \over 3}- \lambda_2\right)
+
{ 4 C_A - 2 N_F \over 6}\,\left({ 8 \over 3}- \lambda_3\right)
\right\}
\nonumber\\&&
+{\cal O}(\epsilon)
\rule{0pt}{15 pt}.
\end{eqnarray}
We set
\begin{equation}
\lambda_1 =2,
\hskip 1 cm 
\lambda_2 = {5\over 3},
\hskip 1 cm 
\lambda_3 = {8\over 3}.
\end{equation}
Then
\begin{equation}
\Delta A_T(q^2) = 0 +{\cal O}(\epsilon).
\end{equation}
Since $\Delta A_T$ is zero, we can add it to $A_T$ to
obtain, after setting $\epsilon$ to zero,
\begin{equation}
A_T(q^2) =
-{ \alpha_s \over 8\pi^2}
\int d\bar q^2\,dx\,d\phi\
{1\over \bar q^2 - q^2}
\sum_{J=0}^2\
{ A_{T,J} \over [\bar q^2/{\cal Q}^2 + 4 x(1-x)]^J},
\label{ATnew}
\end{equation}
where
\begin{eqnarray}
A_{T,0}&=&
-(2 C_A - N_F)\,
{ q^2 + e^2\mu^2 \over \bar q^2 + e^2\mu^2}\,
+ 2 C_A \, x(1-x)\,
{ q^2 + e^{5/3}\mu^2 \over \bar q^2 + e^{5/3}\mu^2}
\nonumber\\
&&+ (4 C_A - 2 N_F)\, x(1-x)\,
{ q^2 + e^{8/3}\mu^2 \over \bar q^2 + e^{8/3}\mu^2}\ ,
\nonumber\\
A_{T,1}&=&
2 C_A \biggl\{ q^2/{\cal Q}^2 + 12 x(1-x)
-{2x(1-x)}
\Bigl[ 
q^2/{\cal Q}^2 + 12 x(1-x)
\Bigr]
\biggr\} ,
\nonumber\\
A_{T,2}&=&
16 C_A\,x(1-x) \biggl\{2  - 8 x(1-x)
+{x(1-x)}
\Bigl[
q^2/{\cal Q}^2 + 8 x(1-x)
\Bigr]
\biggr\} .
\label{ATcoefs}
\end{eqnarray}

\subsection{Timelike components}

We evaluate $A_E(q^2)$ using the integrals in
Table \ref{tab:integraltable} and find
\begin{eqnarray}
A_E(q^2) &=&
-{ \alpha_s \over 8\pi^2} 
\int d\bar q^2\,dx\,d^{1-2\epsilon}\!\phi\
\left[x(1-x)\bar q^2 \over 4\pi^2\tilde\mu^2\right]^{-\epsilon}
{1\over \bar q^2 - q^2}
\sum_{J=0}^2\
{ A^{\prime\prime}_{E,J} \over [\bar q^2/{\cal Q}^2 + 4 x(1-x)]^J}
\nonumber\\
&& - ({\rm pole}),
\label{AEform}
\end{eqnarray}
where
\begin{eqnarray}
A^{\prime\prime}_{E,0}&=&
C_A\biggl\{
-24  x(1-x)
+
(1-\epsilon)\left[
(q^2 - \bar q^2)/{\cal Q}^2 +  1 
- 4 x(1-x)
\right]\biggr\}
+ 4  N_F\, x(1-x),
\nonumber\\
A^{\prime\prime}_{E,1}&=&
8 C_A\,x(1-x)\biggl\{
- 1 + 3 q^2/{\cal Q}^2 + 28 x(1-x)
\biggr\},
\nonumber\\
A^{\prime\prime}_{E,2}&=&
32 C_A\,[x(1-x)]^2 \biggl\{
1 - 3 q^2/{\cal Q}^2 - 16 x(1-x)
 \biggr\}.
\end{eqnarray}
We note that $A^{\prime\prime}_{E,0}$ contains a term
$C_A (1-\epsilon) (q^2 - \bar q^2)/{\cal Q}^2$. This term is
undesirable for numerical integration because it leads to
a quadratically ultraviolet divergent integral. Furthermore, this term
gives a contribution to the integral that vanishes for all $\epsilon$.
For these reasons, we eliminate the term. Then $A_E(q^2)$ still has the
form (\ref{AEform}) but with new coefficients $A^{\prime}_{E,J}$:
\begin{eqnarray}
A^{\prime}_{E,0}&=&
C_A\biggl\{
-24  x(1-x)
+
(1-\epsilon)\left[
 1 
- 4 x(1-x)
\right]\biggr\}
+ 4  N_F\, x(1-x),
\nonumber\\
A^{\prime}_{E,1}&=&
8\,C_A\,x(1-x)\biggl\{
- 1 + 3 q^2/{\cal Q}^2 + 28\, x(1-x)
\biggr\},
\nonumber\\
A^{\prime}_{E,2}&=&
32\,C_A\,[x(1-x)]^2 \biggl\{
1 - 3\,q^2/{\cal Q}^2 - 16\,x(1-x)
 \biggr\}.
\end{eqnarray}

We now examine the ultraviolet renormalization of $A_E(q^2)$. Define
\begin{eqnarray}
\Delta A_E(q^2) &=&
{ \alpha_s \over 8 \pi^2}
\int d\bar q^2\, dx\, d^{1-2\epsilon}\phi
\left[
x(1-x) \bar q^2 \over 4 \pi^2 \tilde \mu^2
\right]^{-\epsilon}
\nonumber\\
&& \times\left\{
(1 - \epsilon)\,
{ C_A
 \over \bar q^2 + e^{\lambda_4}\mu^2}
-
 x(1-x) \,
{ 24 C_A - 4 N_F
 \over \bar q^2 + e^{\lambda_5}\mu^2}
-
 x(1-x)(1-\epsilon) \,
{ 4 C_A
 \over \bar q^2 + e^{\lambda_6}\mu^2}
\right\}
\nonumber\\
&&-
({\rm pole}).
\end{eqnarray}
Here as before $\mu^2 = 4\pi\tilde\mu^2\,e^{-\gamma}$ is the \MSbar\
scale and  $\lambda_4$, $\lambda_5$, and $\lambda_6$ are parameters
that we will adjust. The integrand in $\Delta A_E$ matches that of
$A_E$ when $\bar q^2 \to \infty$ but it has the opposite sign, so if we
add $\Delta A_E$ to $A_E$, we will obtain an ultraviolet convergent
integral. Furthermore, since the integrands match for $\bar q^2 \to
\infty$, the ultraviolet pole terms that are included in the
definitions are opposite and will cancel in the sum $A_E + \Delta
A_E$. We can easily perform the integration:
\begin{eqnarray}
\Delta A_E(q^2) &=&
{ \alpha_s \over 4 \pi}\,
\Gamma(\epsilon)\,
e^{\gamma\epsilon}
\biggl\{
C_A\,
{(1-\epsilon) \Gamma(1-\epsilon)^2 \over \Gamma(2 - 2\epsilon)}\
e^{-\lambda_4\epsilon}
\nonumber\\
&& 
-
(24 C_A - 4 N_F)\,
{ \Gamma(2-\epsilon)^2 \over \Gamma(4 - 2\epsilon)}
e^{-\lambda_5\epsilon}
-
4 C_A\,
{(1-\epsilon) \Gamma(2-\epsilon)^2 \over \Gamma(4 - 2\epsilon)}
e^{-\lambda_6\epsilon}
\biggr\}
- ({\rm pole})
\nonumber\\
&=&
{ \alpha_s \over 4 \pi}
\left\{
 C_A\,
(1-\lambda_4)
-
(24 C_A - 4 N_F)\,
{ 1 \over 6}\,({ 5 \over 3}- \lambda_5)
-
4 C_A \,
{ 1 \over 6}\,({ 2 \over 3}- \lambda_6)
\right\}
\nonumber\\&&
+{\cal O}(\epsilon) .
\rule{0pt}{15 pt}
\end{eqnarray}
We set
\begin{equation}
\lambda_4 = 1,
\hskip 1 cm 
\lambda_5 = {5\over 3},
\hskip 1 cm 
\lambda_6 = {2\over 3}.
\label{lambda456}
\end{equation}
Then
\begin{equation}
\Delta A_E(q^2) = 0 +{\cal O}(\epsilon).
\end{equation}
Since $\Delta A_E$ is zero, we can add it to $A_E$ to
obtain, after setting $\epsilon$ to zero,
\begin{equation}
A_E(q^2) =
-{ \alpha_s \over 8\pi^2}
\int d\bar q^2\,dx\,d\phi\
{1\over \bar q^2 - q^2}
\sum_{J=0}^2\
{ A_{E,J} \over [\bar q^2/{\cal Q}^2 + 4 x(1-x)]^J},
\label{AEnew}
\end{equation}
where
\begin{eqnarray}
A_{E,0}&=&
C_A\
{ q^2 + e^1\mu^2 \over \bar q^2 + e^1\mu^2}\,
- (24 C_A - 4\,N_F)\, x(1-x)\,
{ q^2 + e^{5/3}\mu^2 \over \bar q^2 + e^{5/3}\mu^2}
- 4 C_A\, x(1-x)\,
{ q^2 + e^{2/3}\mu^2 \over \bar q^2 + e^{2/3}\mu^2}\ ,
\nonumber\\
A_{E,1}&=&
8\,C_A\,x(1-x)\biggl\{
- 1 + 3 q^2/{\cal Q}^2 + 28\, x(1-x)
\biggr\},
\nonumber\\
A_{E,2}&=&
32\,C_A\,[x(1-x)]^2 \biggl\{
1 - 3\,q^2/{\cal Q}^2 - 16\,x(1-x)
 \biggr\}.
\end{eqnarray}

\subsection{The tensor $F_g^{\mu\nu}$}

We can assemble this information to obtain $F_g^{\mu\nu}(q)$
defined in Eq.~(\ref{Fgmunu}). With $\vec q$ directed along the $z$
axis, the only non-zero components of $F_g^{\mu\nu}$ are  $F_g^{00}$
and $F_g^{ij}$ for $i,j \in\{1,2\}$: 
\begin{eqnarray}
F_g^{ij}(q)&=&A_T(q^2)\,\delta^{ij}, \hskip 1 cm i,j \in \{1,2\},
\nonumber\\
F_g^{00}(q)&=& { q^2 \over {\cal Q}^2}\,A_E(q^2).
\end{eqnarray}
We can write this in a form valid for any direction of $\vec q$ as
\begin{equation}
F_g^{\mu\nu}(q)=
D(q)^{\mu\nu}
A_T(q^2)
+ n^\mu n^\nu {q^2 \over {\cal Q}^2}\,
\left[
A_E(q^2)
- A_T(q^2)
\right].
\label{Fgtensor0}
\end{equation}
Here $D(q)^{\mu\nu}$ is the numerator for the bare gluon propagator,
Eq.~(\ref{Dmunu}), with $D^{00} = q^2/{\cal Q}^2$. Note that this result
is physically sensible. The complete one loop contribution to the gluon
propagator is $F^{\mu\nu}(q)/q^2$. It has a simple pole times
logarithms at $q^2 = 0$. The pole multiplies a projection onto
transverse polarizations.

\subsection{${W}_g^{\mu\nu}$ and its supplementary terms}

From Eq.~(\ref{Fgtensor0}) we obtain the tensor ${W}_g^{\mu\nu}$ defined
in Eq.~(\ref{Wdef}),
\begin{eqnarray}
{W}_g^{\mu\nu} &=& - { \alpha_s \over 4\pi}\ D(q)^{\mu\nu}
\sum_J { A_{T,J} \over [\bar q^2/{\cal Q}^2 + 4 x(1-x)]^J}
\nonumber\\
&&-{ \alpha_s \over 4\pi}\
n^\mu n^\nu { q^2 \over {\cal Q}^2}
\sum_J {A_{E,J} - A_{T,J} \over [\bar q^2/{\cal Q}^2 + 4 x(1-x)]^J}.
\label{Wstart}
\end{eqnarray}
We can now set $q^2$ to 0 in ${W}_g^{\mu\nu}$ to obtain
${\cal W}_g^{\mu\nu}$. We have
\begin{equation}
{\cal W}_g^{\mu\nu}[{\rm simple}] = - { \alpha_s \over 2\pi}\
D(q)^{\mu\nu}\,
{\cal P}_{\!g}(\bar q^2, x),
\label{calWsimple}
\end{equation}
where
\begin{equation}
{\cal P}_{\!g}(\bar q^2, x) =
{ 1 \over 2} \sum_J { A_{T,J} \over [\bar q^2/{\cal Q}^2 + 4 x(1-x)]^J}.
\end{equation}
Here the coefficients $A_{T,J}$ from Eq.~(\ref{ATcoefs}) are evaluated
with $q^2 = 0$.

We see immediately that there will be a numerical problem for the
cancellation of ${\cal W}_g^{\mu\nu}$ with ${\cal M}_g^{\mu\nu}$ 
at the small $\bar q^2$ endpoint of the integration
(\ref{Igrealandvirt}). In Eq.~(\ref{calMgterms}) for ${\cal
M}_g^{\mu\nu}$, there are terms with four tensor structures. There is a
good cancellation for the $D(q)^{\mu\nu}$ structure (as we will see
below) and the $n^\mu n^\nu$ structure is not important because it
multiplies a factor $\bar q^2$ in  Eq.~(\ref{calMgterms}) for ${\cal
M}_g^{\mu\nu}$. For the other two terms, involving $ l_T^\mu$, the
singularity in ${\cal M}_g^{\mu\nu}$ will vanish if we integrate over
the angle $\phi$ of $ l_T^\mu$ before letting $\bar q^2$ become
small. However, this is not suitable for a numerical integration.
Therefore, we modify ${\cal W}_g^{\mu\nu}$ to
\begin{eqnarray}
{\cal W}_g^{\mu\nu} &=& 
- { \alpha_s \over 2\pi}\ D(q)^{\mu\nu}\,
{\cal P}_{\!g}(\bar q^2, x)
\nonumber\\
&& -{ \alpha_s \over 4\pi} \left[
{  l_T^\mu l_T^\nu \over \vec l_T^{\,2}}
- {\textstyle{1 \over 2}}D(q)^{\mu\nu}
\right]\,
{N_{tt}\over (1 + \bar q^2/{\cal Q}^2)}
\nonumber\\
&&
- { \alpha_s \over 4\pi} { q\cdot n \over (1 + \Delta){\cal Q}^2}
\left( l_T^\mu n^\nu + n^\mu  l_T^\nu \right)\,
{N_{Et}\over (1 + \bar q^2/{\cal Q}^2)}\ ,
\label{calWmod}
\end{eqnarray}
where  $N_{tt}$ and $N_{Et}$ are functions of $\bar q^2$ and $x$ and are
given in Eq.~(\ref{calMcoefs}). The integral of the extra terms
vanishes (because the integration over $\phi$ gives zero), so we are
adding zero to $F_g^{\mu\nu}$. However the cancellation in
Eq.~(\ref{Igrealandvirt}) now works point by point in $\{\bar
q^2,x,\phi\}$ space. We have inserted factors $1/(1 + \bar q^2/{\cal
Q}^2)$ in the extra terms so as not to create problems at $\bar q^2 \to
\infty$ at the same time as we were alleviating problems at $\bar q^2
\to 0$. As in Eq.~(\ref{calMgterms}), we have written ${\cal
W}_g^{\mu\nu}$ in Eq.~(\ref{calWmod}) in a form that is invariant under
the replacements $q^\mu \to - q^\mu,  l^\mu \to -  l^\mu$.

Let us examine the cancellation for $\bar q^2 \to 0$ in
Eq.~(\ref{Igrealandvirt}). The terms with tensor structures involving
$ l_T^\mu$ cancel the corresponding terms in ${\cal M}_g^{\mu\nu}$ by
construction. For the $D(q)^{\mu\nu}$ term, the $\bar q^2 \to 0$ limit
is
\begin{equation}
{\cal W}_g^{\mu\nu} \sim 
- { \alpha_s \over 2\pi}\ D(q)^{\mu\nu}\,
{\cal P}_{\!g}(0,x).
\label{calW0}
\end{equation}
We find for ${\cal P}_{\!g}(\bar q^2, x)$ at $\bar q^2 = 0$,
\begin{equation}
{\cal P}_{\!g}(0,x) =
{\textstyle{1\over 2}}\,\tilde P_{g/g}(x)
+ N_F \tilde P_{q/g}(x),
\label{calPg0}
\end{equation}
where the parton evolution kernels $\tilde P_{g/g}(x)$ and
$\tilde P_{q/g}(x)$ are given in Eq.~(\ref{APkernels}). Using
Eq.~(\ref{calM0}), we see that this is just the behavior we needed to
make the cancellation work.

\section{Structure of graphs with a cut quark propagator}
\label{sec:cutquarkstructure}

In this and the following sections, we analyze the quark
propagator in Coulomb gauge. To set the notation, we write the
contribution from an order $\alpha_s^0$ cut propagator as
\begin{equation}
{\cal I}[{\rm Born}] = \int\!d\vec q\ {\rm Tr}\left\{ 
{ \rlap{/}q \over 2{\cal Q}}\,
R_0
\right\},
\label{quarkborn}
\end{equation}
where, as in Sec.~\ref{sec:cutgluonstructure}, $R_0$ denotes the
factors associated with the rest of the graph and with the final state
measurement function ${\cal S}$. The function  $R_0$ carries hidden
Dirac indices and there is a trace over the Dirac indices of
$\rlap{/}q\,R_0$.

Following the notation employed for the gluon propagator, we write the
contribution of a cut quark self-energy graph as
\begin{equation}
{\cal I}[{\rm real}] = \int\!d\vec q\ {\rm Tr}\left\{ 
{ 1 \over 4 \pi {\cal Q}}\int 
{d\bar q^2\over \bar q^2}\,dx\,d\phi\ 
{\cal M}_q(\bar q^2,x,\phi)\,
R(\bar q^2,x,\phi)
\right\}.
\label{Iqreal}
\end{equation}
With this notation, the contribution from the virtual self-energy
graph with an adjoining cut bare propagator is
\begin{equation}
{\cal I}[{\rm virtual}] = \int\!d\vec q\ {\rm Tr}\left\{ 
{ 1\over 2 {\cal Q}}\,
F_q(q)\,
R_0\right\},
\label{Ivirtq}
\end{equation}
where
\begin{equation}
F_q(q)
={ \rlap{/}q \Sigma(q) \rlap{/}q \over q^2}.
\label{Fquark}
\end{equation}
We should note that ${\cal I}[{\rm virtual}]$ in Eq.~(\ref{Ivirtq}) is
the total contribution from the two graphs in which one or the other of
the neighboring bare propagators is cut. The contribution from either
of these graphs is half of this.

If both of the adjoining bare propagators are uncut, the
corresponding expression is
\begin{equation}
{\cal I}[{\rm all\ uncut}] = \int\!d^4 q\ {\rm Tr}\left\{ 
{ 1\over q^2}\,
F_q(q)\,
R(q^2) \right\},
\end{equation}
where, here, $R$ is defined so that the corresponding Born
contribution is $R(q^2)\rlap{/}q/q^2$.
  
We investigate $F_q(q)$ in the following section. First, we
take $q^2 < 0$. We write $F_q(q)$ as an integral
\begin{equation}
F_q(q) = { 1 \over 2 \pi}\int 
d\bar q^2\,dx\,d\phi\ 
{ 1 \over \bar q^2 - q^2}\,
{W}_q(\bar q^2,x,\phi).
\label{Wqdef}
\end{equation}
Then the contribution to the graph when the adjoining bare
propagators are uncut is
\begin{equation}
{\cal I}[{\rm all\ uncut}] = \int\!d^4 q\ {\rm Tr}\left\{ 
{ 1 \over 2 \pi\, q^2}\int 
d\bar q^2\,dx\,d\phi\ 
{ 1 \over \bar q^2 - q^2}\,
{W}_q(\bar q^2,x,\phi)\,
R(q^2)
\right\}.
\label{Wquse}
\end{equation}

Now, taking $q^2$ to zero, we write $F_q(q)$ as
\begin{equation}
F_q(q) = { 1 \over 2 \pi}\int 
{d\bar q^2\over \bar q^2}\,dx\,d\phi\ 
{\cal W}_q(\bar q^2,x,\phi),
\label{calWqdef}
\end{equation}
where ${\cal W}_{\!q}$ is $W_{\!q}$ with $q^2 = 0$. Then
\begin{eqnarray}
{\cal I}[{\rm real}]+{\cal I}[{\rm virtual}]&=&
\int\!d\vec q\ {\rm Tr}\biggl\{ 
{ 1 \over 4 \pi {\cal Q}}\int {d\bar q^2\over \bar q^2}\,dx\,d\phi\
\label{Iqrealandvirt}\\
&&\times\left[
{\cal M}_q(\bar q^2,x,\phi)\,
R(\bar q^2,x,\phi)
+
{\cal W}_q(\bar q^2,x,\phi)\,
R_0
\right]
\biggr\}.
\nonumber
\end{eqnarray}
As we shall see, the $\bar q^2 \to 0$ singularities in the integrand in
Eq.~(\ref{Iqrealandvirt}) cancel, so that the integral is finite and is
suitable for calculation by numerical integration.

\section{Real quark self-energy graph}
\label{sec:realquarks}

It is straightforward to evaluate the function ${\cal M}_q$
defined in Eq.~(\ref{Iqreal}). We find
\begin{equation}
{\cal M}_q =
 { \alpha_s \over 4\pi}\,{ 1 \over 1 + \Delta}
\left\{N_L\,\rlap{/}q_{\rm os}
+N_{E}\,\Delta\
n\cdot q_{\rm os}\, \rlap{/}n
+ N_{t}\,\rlap{/} l_T
\right\}.
\label{calMqterms}
\end{equation}
The coefficients $N_L,N_{E}$, and $N_{t}$ are functions of
$\bar q^2$ and $x$ and are given below. The momentum in the first
two terms is $q_{\rm os} = (\bar q^0/(1+\Delta),\vec q) = ((k_+^0 +
k_-^0)/(1+\Delta),\vec q)$ so that $q_{\rm os}^2 = 0$. In the third
term, using the notation of Sec.~\ref{sec:elliptical}, $ l_T^\mu$ is
the part of the loop momentum orthogonal to $n^\mu$ and $q_{\rm
os}^\mu$.  Note that this term vanishes if we average over angles
$\phi$.

The coefficients are 
\begin{eqnarray}
N_L &=&
C_F\Biggl\{
12 x(1-x) + (2x-1)(2x+\Delta)
\nonumber\\
&&\ \ \ \ \ 
-{ 16 x(1-x) (2x-1)\over 2x+\Delta}
+{ 16 x(1-x)[1-2x(1-x)] \over (2x+\Delta)^2}
\Biggr\},
\nonumber\\
N_{E} &=&
2 C_F\,{ (1-x)(4 x^2 + \Delta^2) \over (2x+\Delta)^2}\ ,
\nonumber\\
N_{t} &=&
2C_F\left\{
1 - { 2(2x-1) \over 2x+\Delta}
- { 8 x(1-x)\over (2x+\Delta)^2}
\right\}.
\label{calMqcoefs}
\end{eqnarray}

The behavior of ${\cal W}_q^{\mu\nu}$ as $\bar q^2 \to 0$ is important.
Leaving out the $\rlap{/} l_T$ term since $l_T \to 0$ as $\bar q^2 \to
0$, we obtain
\begin{equation}
{\cal M}_q^{\mu\nu} \sim 
\rlap{/}q\,{ \alpha_s \over 2\pi}\
\tilde P_{g/q}(x),
\label{calMq0}
\end{equation}
where
\begin{equation}
\tilde P_{g/q}(x)= C_F\,
{ 1 + (1-x)^2 \over x}
\label{APkernelsbis}
\end{equation}
is the one loop parton evolution kernel for $q \to g$. Notice that
$\tilde P_{q/g}(x)$ is singular at $x \to 0$ but that ${\cal M}_q$
is {\em not} singular at this point. The singularity emerges only
when we take the $\bar q^2 \to 0$ limit.

For the integrand of a cut antiquark self-energy graph, we have ${\cal
M}_{\bar q} = - {\cal M}_{q}$. We note that ${\cal M}_{q}$ in 
Eq.~(\ref{calMqterms}) is odd under the interchange $\bar q^\mu \to - 
\bar q^\mu$, $k_+^\mu \to - k_+^\mu$, $k_-^\mu \to - k_-^\mu$ (so that
also $ l^\mu \to -  l^\mu$). Thus for an antiquark, we can simply
use Eq.~(\ref{calMqterms}) and reverse the momenta, so that $\bar
q^\mu$ flows in the direction of the fermion arrow in the graph. We
use the same principle for the analysis that follows of the virtual
quark self-energy graph.

\section{Virtual quark self-energy graph}
\label{sec:virtualquarks}

In this subsection, we analyze the virtual self-energy graph at
space-like momentum $q^\mu$. We consider the quantity
\begin{equation}
F_q(q) = { \rlap{/}q\, \Sigma(q)\,\rlap{/}q\over q^2}.
\end{equation}
The Feynman rules give
\begin{equation}
F_q(q) =
{ i \over q^2}\,g^2 C_F\,
\tilde\mu^{2\epsilon}
\int { d^{4-2\epsilon}k_+\over (2\pi)^{4-2\epsilon}}\
{ \rlap{/}q\,\gamma_\mu \rlap{/}k_- \gamma_\nu \rlap{/}q\, 
\over (k_+^2 + i\epsilon)(k_-^2 + i\epsilon)}
D(k_+)^{\mu\nu}
-{\rm (pole)}.
\end{equation}

The function $F_q(q)$ must have the form
\begin{equation}
F_q(q) = \rlap{/}q\,B_L(q^2)
+ q\cdot n\,\rlap{/}n\ (q^2 / {\cal Q}^2)\, B_E(q^2).
\label{quarkpropform1}
\end{equation}
The functions $B_L$ and $B_E$ can be extracted using
\begin{eqnarray}
B_L(q^2) &=&
{ -1 \over 4\,{\cal Q}^2}\,{\rm Tr}\left\{F_q(q)
\rlap{/}\tilde q\right\},
\nonumber\\
B_E(q^2)&=&
{ -1 \over 4\, q\cdot n \ q^2}\,{\rm Tr}\left\{F_q(q)
[q^2\,\rlap{/}n - q\cdot n\,\rlap{/} q]\right\}.
\end{eqnarray}

\subsection{Space-like part}

We evaluate $B_L(q^2)$ using the integrals in
Table \ref{tab:integraltable} and dropping terms that are odd under $x
\leftrightarrow (1-x)$, which will integrate to zero. We find
\begin{eqnarray}
B_L(q^2) &=&
-{ \alpha_s \over 8\pi^2}
\int d\bar q^2\,dx\,d^{1-2\epsilon}\!\phi\
\left[x(1-x)\bar q^2 \over 4\pi^2\tilde\mu^2\right]^{-\epsilon}
{1\over \bar q^2 - q^2}
\sum_{J=0}^2\
{ B^{\prime\prime}_{L,J} \over [\bar q^2/{\cal Q}^2 + 4
x(1-x)]^J}
\nonumber\\
&& - ({\rm pole}),
\label{Fqform}
\end{eqnarray}
where
\begin{eqnarray}
B^{\prime\prime}_{L,0}&=&
C_F
\left\{
-4 {{\cal Q}^2 \over q^2} + 24 { {\cal Q}^2 \over q^2}\,x(1-x)
- 1 - \epsilon + 12 x(1-x)
\right\},
\nonumber\\
B^{\prime\prime}_{L,1}&=&
2 C_F
\Biggl\{24 {{\cal Q}^2 \over q^2}\, x(1-x) 
-112 {{\cal Q}^2 \over q^2}\,[x(1-x)]^2
+ 2 + 8 x(1-x)
- 56 [x(1-x)]^2 
\nonumber\\ &&
+{ q^2 \over{\cal Q}^2} - 2 {q^2\over{\cal Q}^2}\, x(1-x)
\Biggr\},
\nonumber\\
B^{\prime\prime}_{L,2}&=&
64 C_F\,[x(1-x)]^2
\left\{
-2{{\cal Q}^2 \over q^2} + 8 {{\cal Q}^2 \over q^2}\, x(1-x)
- 1 +  4 x(1-x)
\right\}.
\label{BLcoefs0}
\end{eqnarray}

We expect that $B_L(q^2)$ is finite at $q^2 = 0$ when $\epsilon$ is
small and negative, but it is not evident from the form above that this
is so. Nevertheless, explicit analytical integration shows that
$q^2\,B_L(q^2) = 0$ at $q^2 = 0$ for any $\epsilon$. In order to get a
form in which the vanishing of $q^2\,B_L(q^2)$  at $q^2 = 0$ is
manifest, we simply subtract $q^2 B_L(q^2)$ at $q^2 = 0$, which is
zero, from $q^2\,B_L(q^2)$. Then $B_L(q^2)$ still has the form
(\ref{Fqform}) but with new coefficients $B^{\prime}_{L,J}$:
\begin{eqnarray}
B^{\prime}_{L,0}&=&
C_F
\{-1 - \epsilon + 12 x(1-x)\},
\nonumber\\
B^{\prime}_{L,1}&=&
2 C_F
\left\{20 x(1-x) - 56 [x(1-x)]^2 
+ {q^2\over{\cal Q}^2} - 2 {q^2\over{\cal Q}^2}\, x(1-x)
\right\},
\nonumber\\
B^{\prime}_{L,2}&=&
32 C_F\,x(1-x)
\left\{1 -  6 x(1-x) + 8 [x(1-x)]^2\right\}.
\label{BLcoefs1}
\end{eqnarray}

We now examine the ultraviolet renormalization of $B_L(q^2)$. We
replace the subtraction of the ultraviolet pole in $4-2\epsilon$
dimensions by a modification of the integrand in four dimensions, just
as we did in the case of the gluon propagator. The result is given by
Eq.~(\ref{Fqform}) with $\epsilon = 0$ and no pole term to subtract,
\begin{equation}
B_L(q^2) =
-{ \alpha_s \over 8\pi^2}
\int d\bar q^2\,dx\,d\phi\
{1\over \bar q^2 - q^2}
\sum_{J=0}^2\
{ B_{L,J} \over [\bar q^2/{\cal Q}^2 + 4
x(1-x)]^J},
\label{Bqfinalform}
\end{equation}
and with new coefficients
$B_{L,J}$:
\begin{eqnarray}
B_{L,0}&=&
C_F
\left\{
-{ q^2 + e^3\mu^2 \over \bar q^2 + e^3\mu^2}
+ 12 x(1-x)\
{ q^2 + e^{5/3}\mu^2 \over \bar q^2 + e^{5/3}\mu^2}
\right\},
\nonumber\\
B_{L,1}&=&
2 C_F
\left\{20 x(1-x) - 56 [x(1-x)]^2 
+ {q^2\over{\cal Q}^2}[1 - 2  x(1-x)]
\right\},
\nonumber\\
B_{L,2}&=&
32 C_F\,x(1-x)
\left\{1 -  6 x(1-x) + 8 [x(1-x)]^2\right\}.
\label{BLcoefs}
\end{eqnarray}

\subsection{Timelike part}

We evaluate $B_E(q^2)$ using the integrals in Table
\ref{tab:integraltable} and dropping terms that are odd under $x
\leftrightarrow (1-x)$, which will integrate to zero. We find
\begin{eqnarray}
B_E(q^2) &=&
-{ \alpha_s \over 8\pi^2}
\int d\bar q^2\,dx\,d^{1-2\epsilon}\!\phi\
\left[x(1-x)\bar q^2 \over 4\pi^2\tilde\mu^2\right]^{-\epsilon}
{1\over \bar q^2 - q^2}
\sum_{J=0}^2\
{ B^{\prime}_{E,J} \over [\bar q^2/{\cal Q}^2 + 4 x(1-x)]^J}
\nonumber\\
&& - ({\rm pole}),
\label{Fnform}
\end{eqnarray}
where
\begin{eqnarray}
B^{\prime}_{E,0}&=&
2 C_F\,{ {\cal Q}^2 \over q^2}\,
\{1 -  6 x(1-x)\},
\nonumber\\
B^{\prime}_{E,1}&=&
2 C_F
\left\{- 12 x(1-x){ {\cal Q}^2 \over q^2} 
+ 56 [x(1-x)]^2\, { {\cal Q}^2 \over q^2}
- 1 + 2 x(1-x)
\right\},
\nonumber\\
B^{\prime}_{E,2}&=&
64 C_F\,[x(1-x)]^2\,{ {\cal Q}^2 \over q^2}\,
\{1 -  4 x(1-x)\}.
\label{BEcoefs0}
\end{eqnarray}
We expect that $B_E(q^2)$ is finite at $q^2 = 0$ when $\epsilon$ is
small and negative. It is, however, not evident from the form above that
this is so. Nevertheless, explicit analytical integration shows that
$q^2 B_E(q^2)$ vanishes at $q^2 = 0$ for any $\epsilon$. In order to
get a form in which this vanishing is manifest, we simply subtract
$q^2 B_E(q^2)$ at $q^2 = 0$, which is zero, from $q^2 B_E(q^2)$. Then
$B_E(q^2)$ still has the form (\ref{Fqform}) but with new coefficients
$B_{E,J}$:
\begin{eqnarray}
B_{E,0}&=&0,
\nonumber\\
B_{E,1}&=&
- 8 C_F\,x(1-x),
\nonumber\\
B_{E,2}&=&
16 C_F\,x(1-x)
\{-1 +  4 x(1-x)\}.
\label{BEcoefs}
\end{eqnarray}

The pole term vanishes, so we can simply set $\epsilon$ to zero,
obtaining
\begin{equation}
B_E(q^2) =
-{ \alpha_s \over 8\pi^2}
\int d\bar q^2\,dx\,d\phi\
{1\over \bar q^2 - q^2}
\sum_{J=0}^2\
{ B_{E,J} \over [\bar q^2/{\cal Q}^2 + 4 x(1-x)]^J}
\label{newBnform}
\end{equation}
with the same coefficients (\ref{BEcoefs}).

\subsection{$W_q$ and its supplementary terms}

We can summarize our results so far by writing 
\begin{equation}
F_q(q) = { 1 \over 2 \pi}\int 
{d\bar q^2\over \bar q^2-q^2}\,dx\,d\phi\ 
{W}_q(\bar q^2,x,\phi),
\label{Wqagain}
\end{equation}
where $W_q$ has the form
\begin{equation}
W_q = -{\alpha_s \over 4\pi}\left\{
\rlap{/}q\,
U_L(x,\bar q^2,q^2) 
+{ q^2 \over {\cal Q}^2}\, q\cdot n\,\rlap{/}n\,
U_E(x,\bar q^2,q^2)
\right\}.
\label{WfromULandUE}
\end{equation}
with
\begin{eqnarray}
U_L(x,\bar q^2,q^2) &=&
\sum_{J=0}^2{ B_{L,J} \over [\bar q^2/{\cal Q}^2 + 4 x(1-x)]^J}\ ,
\nonumber\\
U_E(x,\bar q^2,q^2)&=&
\sum_{J=0}^2{ B_{E,J} \over [\bar q^2/{\cal Q}^2 + 4 x(1-x)]^J}.
\end{eqnarray}
The coefficients $B_{L,J}$ and $B_{E,J}$ are given in
Eqs.~(\ref{BLcoefs}) and (\ref{BEcoefs}). Then the contribution to the
full graph from the quark self-energy graph when the adjoining bare
propagators are uncut is given in terms of $W_q$ by Eq.~(\ref{Wquse}).

We can now set $q^2$ to 0 in ${W}_q$ to obtain ${\cal W}_q$. The
second term in Eq.~(\ref{WfromULandUE}) vanishes and we have
\begin{equation}
{\cal W}_q[{\rm simple}] = -{\alpha_s \over 4\pi}\,
\rlap{/}q\,U_L(x,\bar q^2,0).
\end{equation}
We see immediately that there will be a numerical problem for the
cancellation of ${\cal W}_q$ with ${\cal M}_q$ at the small $\bar q^2$
endpoint of the integration (\ref{Iqrealandvirt}). In
Eq.~(\ref{calMqterms}) for ${\cal M}_q$, there are terms with three
Dirac matrix structures.  The $\rlap{/}n$ term is not important because
it multiplies a factor $\Delta$ and $\Delta \propto
\bar q^2$ for small $\bar q^2$. In the term proportional to
$\rlap{/} l_T$, the singularity in ${\cal M}_g^{\mu\nu}/\bar q^2$ will
vanish if we integrate over the angle $\phi$ of $ l_T^\mu$ before
letting
$\bar q^2$ become small. However, this is not suitable for a numerical
integration. Finally, the coefficient of $\rlap{/}q$ in ${\cal W}_q$
contains only terms that are even under $x \to (1-x)$, while the
coefficient of $\rlap{/}q$ in ${\cal M}_q$ contains both even and odd
terms. The small $\bar q^2$ singularity will be cancelled if we
integrate over $x$ before letting $\bar q^2$ become small. But, again,
this is not suitable for a numerical integration. 

We can make the
cancellation happen point by point in $x$ and $\phi$ by  adding two
terms to ${\cal W}_q$, so that it becomes
\begin{equation}
{\cal W}_q =
-{\alpha_s \over 4\pi}\Biggl\{
\rlap{/}q\,
U_L(x,\bar q^2,0)
+\rlap{/}q\,
{ V_L(x,\bar q^2) \over (1+\bar q^2/{\cal Q}^2)}
+\rlap{/} l_T\,
{ N_t(x,\bar q^2) \over (1+\bar q^2/{\cal Q}^2)}
\Biggr\}. 
\end{equation}
Here the function $V_L(x,\bar q^2)$ is related to the coefficient
$N_L(x,\bar q^2)$ of $\rlap{/} q$ in the cut self-energy,
Eq.~(\ref{calMqterms}), by
\begin{equation}
V_L(x,\bar q^2) =
{ 1 \over 2(1 + \Delta)}\left[
N_L(x,\bar q^2) - N_L(1-x,\bar q^2)
\right].
\label{VLdef}
\end{equation}
(The factor $1/(1+\Delta)$ here makes the calculated expression for
$V_L$ simpler.) The function $N_t(x,\bar q^2)$ is the coefficient of
$\rlap{/}  l$ in the cut self-energy, Eq.~(\ref{calMqterms}), and is
given in Eq.~(\ref{calMqcoefs}). 

The integrals of the extra terms vanish (because the integrations
over $x$ and $\phi$ respectively give zero), so we are adding zero to
${\cal W}_q$. However the cancellation in Eq.~(\ref{Iqrealandvirt})
can now work point by point in $\{\bar q^2,x,\phi\}$ space. (We
check this for the $\rlap{/}q$ terms below.) We have inserted factors
$1/(1 + \bar q^2/{\cal Q}^2)$ in the extra terms so as not to create
problems at $\bar q^2 \to \infty$ at the same time as we were
alleviating problems at $\bar q^2 \to 0$.

Let us summarize. The expression for ${\cal W}_q$ in
Eq.~(\ref{calWqdef}) is now
\begin{equation}
{\cal W}_q =
-{\alpha_s \over 2\pi}\,
\rlap{/}q\,
{\cal P}_{\!q}(\bar q^2, x)
-{\alpha_s \over 4\pi}\rlap{/} l_T\,
{ N_t(x,\bar q^2) \over (1+\bar q^2/{\cal Q}^2)}, 
\end{equation}
with
\begin{eqnarray}
{\cal P}_{\!q}(\bar q^2, x) &=&
{ 1 \over 2}
\left\{  U_L(x,\bar q^2,0)
+
{ V_L(x,\bar q^2) \over (1+\bar q^2/{\cal Q}^2)}\right\}
\nonumber\\
&=& 
{ 1 \over 2}
\sum_{J=0}^2{ B_{L,J} \over [\bar q^2/{\cal Q}^2 + 4 x(1-x)]^J}
\nonumber\\
&&
+ { 1 \over 2}\,{ 1 \over (1+\bar q^2/{\cal Q}^2)}
\sum_{J=0}^2{ N_{L,J} \over [\bar q^2/{\cal Q}^2 + 4 x(1-x)]^J}\ ,
\nonumber\\
N_t(x,\bar q^2) &=&
\sum_{J=0}^2{ N_{t,J} \over (\Delta + 2 x)^J}.
\end{eqnarray}
Notice that the denominators in the $\rlap{/} l_T$ term are different
from those in the $\rlap{/}q$ terms. The coefficients
$B_{L,J}$ from Eq.~(\ref{BLcoefs}) are here evaluated at $q^2 = 0$:
\begin{eqnarray}
B_{L,0}&=&
C_F
\left\{
-{ e^3\mu^2 \over \bar q^2 + e^3\mu^2}
+ 12 x(1-x)\
{e^{5/3}\mu^2 \over \bar q^2 + e^{5/3}\mu^2}
\right\},
\nonumber\\
B_{L,1}&=&
8 C_F x(1-x)
\left\{5  - 14 x(1-x) 
\right\},
\nonumber\\
B_{L,2}&=&
32 C_F\,x(1-x)
\left\{1 -  6 x(1-x) + 8 [x(1-x)]^2\right\}.
\label{Bqn0}
\end{eqnarray}
The coefficients $N_{L,J}$ are computed from the coefficients for
$N_L(x, \bar q^2)$ in {Eq.~(\ref{calMqcoefs})} and are
\begin{eqnarray}
N_{L,0}&=&
C_F(2x-1) ,
\nonumber\\
N_{L,1}&=&
-16 C_F (2x-1) x(1-x),
\nonumber\\
N_{L,2}&=&
-32 C_F (2x-1)  x(1-x) [1 - 2 x(1-x)].
\label{NLcoefs}
\end{eqnarray}
The coefficients $N_{t,J}$ for $N_t(x,\bar q^2)$ are given in
Eq.~(\ref{calMqcoefs}):
\begin{eqnarray}
N_{t,0}&=&
2 C_F,
\nonumber\\
N_{t,1}&=&
-4 C_F (2x-1),
\nonumber\\
N_{t,2}&=&
-16 C_F x(1-x).
\label{Ntcoefs}
\end{eqnarray}

If we take the $\bar q^2 \to 0$ limit of ${ \cal W}_q$, the 
$\rlap{/} l_T$ term does not contribute since $ l_T \to 0$ as
$\bar q^2 \to 0$. We are left with
\begin{equation}
{\cal W}_q \sim
-{\alpha_s \over 2\pi}\,
\rlap{/}q\,
{\cal P}_{\!q}(0,x). 
\end{equation}
Evaluating ${\cal P}_{\!q}(\bar q^2, x)$ at $\bar q^2 = 0$ we find
\begin{equation}
{\cal P}_{\!q}(0,x) =
\tilde P_{g/q}(x),
\label{calPqlimit}
\end{equation}
where $\tilde P_{g/q}(x)$ is the parton evolution kernel given in
Eq.~(\ref{APkernelsbis}). Thus ${\cal W}_q$ properly cancels ${\cal
M}_q$ in Eq.~(\ref{Iqrealandvirt}).

\section{Renormalization of three point functions}
\label{sec:3pointrenormalization}

In this section, we consider how to renormalize the divergent one loop
virtual three point functions in Coulomb gauge using numerical
integration. 

\subsection{Quark-antiquark-boson vertices}

In this subsection, we construct the renormalization counter term as an
integral over the four dimensional space of loop momenta. We begin
with the corresponding integrals over a $4-2\epsilon$ dimensional
space, since we want to match the renormalization to standard \MSbar\
renormalization. We first study the quark-antiquark-gluon vertex. Then
we extend the result to the quark-antiquark-photon vertex, which has a
somewhat simpler structure.

There are two contributions to the quark-antiquark-gluon vertex
$\Gamma_a^\mu(k_1,k_2)$ at one loop. Each of them has the form
\begin{equation}
\Gamma_a^\mu(k_1,k_2) =
ig^2\,C t_a\,
\tilde \mu^{2\epsilon}\!\!
\int\! {d^{4-2\epsilon} l\over(2\pi)^{4-2\epsilon}}\,
{ N( l,k_1,k_2)^\mu_\nu\,\gamma^\nu \over 
[(k_2- l)^2 + i\epsilon]
[(k_1- l)^2 + i\epsilon]
[ l^2 + i\epsilon]},
\end{equation}
where $C$ is the color factor for that graph and the scale factor
$\tilde \mu$ is related to the \MSbar\ scale factor $\mu$ by $\mu^2 =
4\pi\tilde\mu^2 e^{-\gamma}$. When the loop momentum is large, the
function $N$ has the form
\begin{eqnarray}
N^\mu_\nu &=& 
  A_1(\epsilon)\,  l^2 g^\mu_\nu
+ A_2(\epsilon)\, { l^2 \over \tilde  l^2}\  l^2 \,g^\mu_\nu
\nonumber\\
&&
+ A_3(\epsilon)\,  l^\mu  l_\nu
+ A_4(\epsilon)\,{ l^2 \over \tilde  l^2}\ 
(  l^\mu\tilde l_\nu
 + \tilde l^\mu  l_\nu -  l^\mu l_\nu)
\nonumber\\
&&
+ A_5(\epsilon)\, { l^2 \over \tilde  l^2}\
( l^\mu\tilde l_\nu
- \tilde l^\mu  l_\nu)
+ A_6(\epsilon)\,\left( l^2 \over \tilde  l^2\right)^2\ 
\tilde l^\mu( l_\nu - \tilde l_\nu)
\nonumber\\
&&
+{\cal O}( l).
\label{tildeN}
\end{eqnarray}
Here the omitted terms are suppressed by one or more powers of
$k_1/ l$ or $k_2/ l$.

We subtract a suitably chosen quantity $\tilde\Gamma_a^\mu$ from
$\Gamma_a^\mu(k_1,k_2)$:
\begin{equation}
\tilde\Gamma_a^\mu =
ig^2\,Ct_a\,
\tilde \mu^{2\epsilon}\!\!
\int\! {d^{4-2\epsilon} l\over(2\pi)^{4-2\epsilon}}\,
C( l)^\mu_\nu\,\gamma^\nu .
\label{tildegamma}
\end{equation}
Using $D( l^2) =  l^2 - M^2 + i\epsilon$ and $n = (1,0,0,0)$ we
define
\begin{eqnarray}
C^\mu_\nu &=& 
  { A_1(\epsilon) \over D( l^2)^2}\ g^\mu_\nu
+ {A_2(\epsilon) \over D( l^2)\, D(\tilde  l^2)}\ g^\mu_\nu
\nonumber\\
&&
+{ A_3(\epsilon) \over D( l^2)^3}\   l^\mu  l_\nu
+{ A_4(\epsilon) \over  D( l^2)^2\,D(\tilde  l^2)}\ 
(  l^\mu\tilde l_\nu + \tilde l^\mu  l_\nu -  l^\mu l_\nu)
\nonumber\\
&&
+ { A_5(\epsilon) \over  D( l^2)^2\,D(\tilde  l^2)}\ 
( l^\mu\tilde l_\nu
- \tilde l^\mu  l_\nu)
+ {A_6(\epsilon) \over  D( l^2)\,D(\tilde  l^2)^2 }\ 
\tilde l^\mu( l_\nu - \tilde l_\nu)
\nonumber\\
&&
+{ B_1(\epsilon)\,M^2 \over D( l^2)^3}\ g^\mu_\nu
+{ B_2(\epsilon)\,M^2 \over D( l^2)^3}\ n^\mu n_\nu.
\label{Cmunu}
\end{eqnarray}
The functions $B_1(\epsilon)$ and $B_2(\epsilon)$ are defined in terms
of the functions $A_i(\epsilon)$ in such a way that $\tilde \Gamma$ has
a very simple dependence on $\epsilon$. We will give the definition
below.

Two properties of $\tilde \Gamma_a^\mu$ are important. First, the
leading $( l)^{-4}$ behavior of the integrand of $\tilde
\Gamma_a^\mu$ matches that for $\Gamma_a^\mu$ for large $ l^\mu$. 
Second, it is easy to compute $\tilde \Gamma_a^\mu$. 

To begin the computation, we recognize that the integral of various
terms in $C^\mu_\nu$ will be proportional to combinations of
$g^\mu_\nu$ and $n^\mu n_\nu$. This allows us to replace the integrand
by
\begin{eqnarray}
C^\mu_\nu &\to& 
  { A_1(\epsilon) \over D( l^2)^2}\ g^\mu_\nu
+ {A_2(\epsilon) \over D( l^2)\, D(\tilde  l^2)}\ g^\mu_\nu
+{ A_3(\epsilon) \over (4-2\epsilon) }\  g^\mu_\nu\
\left({1 \over D( l^2)^2} 
+{M^2 \over D( l^2)^3}
\right)
\nonumber\\
&&
+ A_4(\epsilon) \ 
{ g^\mu_\nu +(2-2\epsilon) n^\mu n_\nu \over (3-2\epsilon)}
\left(
{ 1 \over D( l^2)^2}
+{M^2 \over D( l^2)^2\,D(\tilde l^2)}
\right) 
\nonumber\\
&&
- A_4(\epsilon) \ 
n^\mu n_\nu
\left(
{1 \over  D( l^2)\,D(\tilde  l^2)}
+{M^2 \over  D( l^2)^2\,D(\tilde  l^2)}
\right)
\nonumber\\
&&
+{ B_1(\epsilon)\,M^2 \over D( l^2)^3}\ g^\mu_\nu
+{ B_2(\epsilon)\,M^2 \over D( l^2)^3}\ n^\mu n_\nu.
\end{eqnarray}
Note that the contributions from $A_5$ and $A_6$ vanish after
integration because the tensors that multiply them in Eq.~(\ref{Cmunu})
vanish when contracted with $g_\mu^\nu$ or $n_\mu n^\nu$.

Now we can perform the integration using
\begin{equation}
I_{J,K} \equiv i\, \tilde \mu^{2 \epsilon} 
\int { d^{4-2\epsilon} l \over (2 \pi)^{4-2\epsilon}}
{
(M^2)^{J+K-2} \over 
[ l^2 - M^2 + i\epsilon]^J [\tilde  l^2 - M^2]^K  }
= \lambda(J,K,\epsilon)\,{ \Gamma(\epsilon) \over 16 \pi^2}
\left(M^2 \over 4\pi \tilde\mu^2\right)^{-\epsilon},
\end{equation}
where
\begin{equation}
\lambda(J,K,\epsilon) =
(-1)^{J+K+1}
{ \Gamma(J-1/2)\,\Gamma(J+K-2+\epsilon) 
\over \Gamma(J)\,\Gamma(J+K-1/2)\,\Gamma(\epsilon)}.
\end{equation}
Specifically
\begin{equation}
\matrix{
\lambda(2,0,\epsilon) = -1 ,
&\lambda(1,1,\epsilon) =-2 ,\cr
\cr
\lambda(3,0,\epsilon) = \epsilon/2 ,
&\lambda(2,1,\epsilon)=2\epsilon/3 . }
\end{equation}
We find
\begin{equation}
\tilde\Gamma_a^\mu = { \alpha_s \over 4\pi}\,Ct_a\,G^\mu_\nu\,\gamma^\nu
\Gamma(\epsilon) 
\left(M^2 \over 4\pi \tilde\mu^2\right)^{-\epsilon}
\label{renormvrtx}
\end{equation}
with (after taking cancellations among the $\epsilon$ dependent factors
into account)
\begin{eqnarray}
G^\mu_\nu &=& 
 - g^\mu_\nu\left\{
A_1(\epsilon) + 2 A_2(\epsilon) + {1 \over 4}\,A_3(\epsilon)
+ {1 \over 3}\,A_4(\epsilon) - {\epsilon \over 2}\,B_1(\epsilon)
\right\}
\nonumber\\
&&
+ n^\mu n_\nu\left\{
{4 \over 3}\,A_4(\epsilon) + {\epsilon \over 2}\,B_2(\epsilon)
\right\}.
\end{eqnarray}
We set
\begin{eqnarray}
B_1(\epsilon)&=&{2 \over \epsilon}
\left\{
[A_1(\epsilon) - A_1(0)] 
+ 2 [A_2(\epsilon) - A_2(0)] 
+ {1 \over 4}\,[A_3(\epsilon) - A_3(0)]
+ {1 \over 3}\,[A_4(\epsilon) - A_4(0)]
\right\},
\nonumber\\
B_2(\epsilon)&=&-{8 \over 3\epsilon}
[A_4(\epsilon) - A_4(0)].
\end{eqnarray}
Then $G^\mu_\nu$ takes the $\epsilon$-independent form
\begin{equation}
G^\mu_\nu = 
 - g^\mu_\nu\left\{
A_1(0) + 2 A_2(0) + {1 \over 4}\,A_3(0)
+ {1 \over 3}\,A_4(0) 
\right\}
+ n^\mu n_\nu\
{4 \over 3}\,A_4(0).
\end{equation}

Expanding Eq.~(\ref{renormvrtx}) about $\epsilon = 0$, we have
\begin{equation}
\tilde\Gamma_a^\mu =
\left[\tilde\Gamma_a^\mu\right]_{\rm pole}
+
\left[\tilde\Gamma_a^\mu\right]_{\rm R} + {\cal O}(\epsilon),
\end{equation}
with
\begin{eqnarray}
\left[\tilde\Gamma_a^\mu\right]_{\rm pole} &=&  
{\alpha_s \over 4\pi}\,C t_a\, 
G^\mu_\nu\, \gamma^\nu\,
{ 1 \over \epsilon},
\nonumber\\
\left[\tilde\Gamma_a^\mu\right]_{\rm R} &=&
{\alpha_s \over 4\pi}\,C t_a\, 
G^\mu_\nu\, \gamma^\nu\, 
\ln\!\left({\mu^2 \over M^2}\right) .
\end{eqnarray}
Define $[\Gamma_a^\mu]_R$ by
\begin{equation}
\Gamma_a^\mu =
\left[\Gamma_a^\mu\right]_{\rm pole}
+
\left[\Gamma_a^\mu\right]_{\rm R} + {\cal O}(\epsilon).
\end{equation}
We recognize that
\begin{equation}
\left[\Gamma_a^\mu\right]_{\rm pole} =
\left[\tilde\Gamma_a^\mu\right]_{\rm pole}
\end{equation}
because the $ l^{-4}$ behaviors of the two integrands match. Thus
we can use our results for $\tilde\Gamma_a^\mu$ to write
\begin{equation}
\left[\Gamma_a^\mu\right]_{\rm R}
=
\left\{
\Gamma_a^\mu - \tilde\Gamma_a^\mu
\right\}_{\epsilon = 0} 
+ \left[\tilde\Gamma_a^\mu\right]_{\rm R}.
\label{renormalize3}
\end{equation}
The first term can be evaluated by numerical integration in 4
dimensions. The second term vanishes if we set
\begin{equation}
M^2 = \mu^2.
\end{equation}

Now we need the coefficients $A_J(\epsilon)$ in Eqs.~(\ref{tildeN}) and
(\ref{Cmunu}). For the graph in which the gluon connects to the gluon
line, one finds
\begin{eqnarray}
&&C= C_A/2,
\nonumber\\
&&A_1(\epsilon) = 2, \ \ \ \ \
A_2(\epsilon) = -2, \ \
A_3(\epsilon) = -4(1-\epsilon),
\nonumber\\
&& A_4(\epsilon) = -2, \ \ \
A_5(\epsilon) = 0, \ \ \ \ 
A_6(\epsilon) = 2,
\nonumber\\
&& B_1(\epsilon) = 2, \ \ \ \ \
B_2(\epsilon) = 0.
\label{Gammacoefs1}
\end{eqnarray}
Here the values for the $B_i(\epsilon)$ have been calculated from the
values for the $A_i(\epsilon)$. For the graph in which the gluon
connects to the quark line, one finds
\begin{eqnarray}
&&C = C_F - C_A/2 =-1/(2N_C),
\nonumber\\
&&A_1(\epsilon) = 2\epsilon, \ \
A_2(\epsilon) = -1, \ \
A_3(\epsilon) = 4(1-\epsilon),
\nonumber\\
&&A_4(\epsilon) = 0, \ \ \
A_5(\epsilon) = 2, \ \ \ \ \,
A_6(\epsilon) = 0,
\nonumber\\
&& B_1(\epsilon) = 2, \ \ \
B_2(\epsilon) = 0.
\label{Gammacoefs2}
\end{eqnarray}
If we change $C$ from $C_F - C_A/2$ to $C_F$, this
same result holds for the quark-antiquark-photon vertex, with the
appropriate change in the color structure from $C t_a$ to $C$.

\subsection{Performing the energy integrations}

We have seen how to renormalize the virtual three point functions in a
fashion that works in four dimensions but is equivalent to \MSbar\
renormalization. Here, we deal with the implementation of the loop
integrals as numerical integrals. We need to write the counter
terms for the three point function as integrals over the space
components of the loop momentum:
\begin{equation}
\tilde\Gamma_a^\mu =
g^2\,t_a\,
\int\! {d^{3}\vec l\over(2\pi)^{3}}\,
E(\vec l)^\mu_\nu\,\gamma^\nu.
\label{tildegamma3}
\end{equation}
To do this, we perform the integrations over the loop energy
analytically. Let us define the integrals
\begin{equation}
I_{N,J} = i\, \omega^{2N-J-1} \int\! { d l^0 \over 2\pi}\
{( l^0)^J \over D( l^2)^N}\ ,
\end{equation}
where
\begin{equation}
\omega = \sqrt{\vec l^2 + M^2}.
\end{equation}
Then
\begin{eqnarray}
E^\mu_\nu &=& 
  {C A_1(0) \over \omega^3}\ g^\mu_\nu\,I_{2,0}
- {C A_2(0) \over \omega^3}\ g^\mu_\nu\,I_{1,0}
\nonumber\\
&&
+{C A_3(0) \over\omega^5}
  \left\{ \tilde  l^\mu \tilde l_\nu\, I_{3,0}
          +\omega^2 n^\mu n_\nu\, I_{3,2}\right\}
-{C A_4(0) \over  \omega^5}\ 
 \left\{\tilde  l^\mu\tilde l_\nu\, I_{2,0}
-\omega^2 n^\mu n_\nu\, I_{2,2}
 \right\}
\nonumber\\
&&
+{C B_1(0)\,M^2 \over \omega^5}\ g^\mu_\nu\, I_{3,0}
+{C B_2(0)\,M^2 \over  \omega^5}\ n^\mu n_\nu\, I_{3,0}.
\end{eqnarray}
Here we have used the fact that the $I_{N,J}$ for odd $J$ vanish.

The integrands of $I_{N,J}$ have singular factors of the form
\begin{equation}
{ 1 \over D( l^2)^N} \equiv
{ 1 \over ( l^2 - M^2 + i\epsilon)^N}
= 
{ 1 \over ( l^0 - \omega + i\epsilon)^N}\
{ 1 \over ( l^0 + \omega - i\epsilon)^N}.
\end{equation}
To perform the integrals, we use the prescription in
Sec.~\ref{sec:energyintegrals}. We close the integration contour in the
lower half plane and then in the upper half plane and take the average
of the results. We get $I_{N,1}=0$ and
\begin{eqnarray}
&I_{1,0} = { 1 / 2},\ \ \ \  &I_{1,2} = { 1/ 2},
\nonumber\\
&\,I_{2,0} = -{ 1 /4}, \ \ \   &I_{2,2}= { 1/ 4},
\nonumber\\
&I_{3,0} = { 3 / 16},\ \ \   &I_{3,2}= -{ 1/ 16}.
\end{eqnarray}
Thus
\begin{eqnarray}
E^\mu_\nu &=& 
- {C A_1(0) \over 4\omega^3}\ g^\mu_\nu
- {C A_2(0) \over 2 \omega^3}\ g^\mu_\nu
\nonumber\\
&&
+{C A_3(0) \over 16\omega^5}
  \left\{3 \tilde  l^\mu \tilde l_\nu
          - \omega^2 n^\mu n_\nu\right\}
+{C A_4(0) \over  4\omega^5}\ 
 \left\{\tilde  l^\mu\tilde l_\nu
+\omega^2 n^\mu n_\nu
 \right\}
\nonumber\\
&&
+{3C  B_1(0)\,M^2 \over 16\omega^5}\ g^\mu_\nu
+{3C  B_2(0)\,M^2 \over 16 \omega^5}\ n^\mu n_\nu.
\end{eqnarray}

We can use our explicit results for the coefficients $A_J$ and $B_J$ to
obtain the net result for the counter term for the
quark-antiquark-gluon one loop graphs (summed over the two graphs),
expressed as an integral over $ \vec l$. The counter term is given by
Eq.~(\ref{tildegamma3}) with
\begin{equation}
E^\mu_\nu =
{C_F \over 8\omega^5}\,(4\omega^2 + 3 M^2)\, g^\mu_\nu
+{3\,C_F - 4\,C_A \over 4\omega^5}\
\tilde  l^\mu \tilde l_\nu
-{C_F \over  4\omega^3}\ n^\mu n_\nu.
\end{equation}
According to Eq.~(\ref {renormalize3}), we are to subtract $\tilde
\Gamma_a^\mu$ from the integral for $\Gamma_a^\mu$ and set $M$ = $\mu$.

For the quark-antiquark-photon graph the counter term is given by
Eq.~(\ref{tildegamma3}) without the factor of $t_a$ and with
\begin{equation}
E^\mu_\nu =
{C_F \over 8\omega^5}\,(4\omega^2 + 3 M^2)\, g^\mu_\nu
+{3\,C_F\over 4\omega^5}\
\tilde  l^\mu \tilde l_\nu
-{C_F \over  4\omega^3}\ n^\mu n_\nu.
\end{equation}

\section{Results}
\label{sec:results}

\begin{figure}
\includegraphics[width = 4 cm]{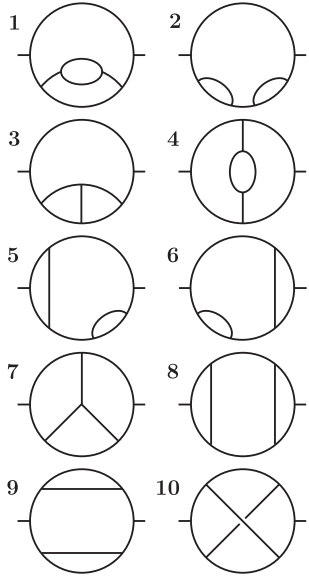}
\medskip
\caption{The ten topologies of Feynman diagrams that contribute order
$\alpha_s^2$ terms to $e^+ e^- \to {\it hadrons}$. The incoming and
outgoing lines are electroweak vector bosons. The other lines can
represent either quarks or gluons. Then a particular contribution to
the cross section is given by a particular cut of the diagram, as in
Fig.~\ref{fig:cutdiagrams}.}
\label{fig:thegraphs}
\end{figure}

In this section, we look at some numerical results. As our example, we
will consider one of the standard event shape variables, the thrust
$t$. We examine the thrust distribution normalized to the
total cross section $\sigma \approx (1 + \alpha_s/\pi)\times \sigma_0$
for $e^+ + e^- \to {\it hadrons}$,
\begin{equation}
{\cal I}(t) = { 1 \over \sigma}\,{ d\sigma \over dt}.
\end{equation}
For $t<1$, $d\sigma/dt$ has a contribution, $d\sigma^{(1)}/dt$, of
order $\alpha_s^1$ and a contribution, $d\sigma^{(2)}/dt$, of
order $\alpha_s^2$. Both contributions are included in the
next-to-leading order results for ${\cal I}(t)$. We also isolate the
second order term, $d\sigma^{(2)}/dt$ and study the second order
contributions to the moments of the thrust distribution,
\begin{equation}
{I}_n = { 1 \over \sigma_0 (\alpha_s/\pi)^2}
\int_0^1\!dt\ (1-t)^n\ {
d\sigma^{(2)} \over dt}.
\label{thrustdef}
\end{equation}

The first question is whether the gauge choice makes any difference.
There are ten topologies of Feynman diagrams that contribute order
$\alpha_s^2$ terms to $e^+ e^- \to {\it hadrons}$. These are shown in
Fig.~\ref{fig:thegraphs}. For each topology, we calculate the
corresponding contribution to the second moment of the thrust
distribution, $I_2$. The results are shown in Table~\ref{tab:diagrams}.
We see that graph by graph, the results are completely different in
Feynman and Coulomb gauges. However, the total $I_2$ summed over graphs
is independent of the gauge.

\begin{table}[htb]
\caption{\label{tab:diagrams}
Comparison of results in Feynman gauge and Coulomb gauge for
$I_2$, the second moment of the thrust distribution,
Eq.~(\ref{thrustdef}). The results are shown for each of the ten graph
topologies in Fig.~\ref{fig:thegraphs}. The errors are not shown,
but are about 1\%. The renormalization scale $\mu$ is chosen to be $\mu
=\sqrt s$.}
\begin{ruledtabular}
\begin{tabular}{ldd}
\multicolumn{1}{c}{graph}
&\multicolumn{1}{c}{Feynman gauge}
&\multicolumn{1}{c}{Coulomb gauge}\\
\colrule
 1    &      0.1125      &    -0.3274      \\
 2    &     -0.03154    &     -0.01422   \\
 3    &      0.2083      &     1.031       \\
 4    &      0.1230      &    -0.1955      \\
 5    &     -0.1602      &    -0.1834      \\
 6    &     -0.1597      &    -0.1828      \\
 7    &      1.820       &     1.584       \\
 8    &     -0.2222      &    -0.1409      \\
 9    &     -0.01243   &       0.03432    \\
 10   &     -0.1227      &    -0.04512    \\
 TOTAL  &    1.555       &     1.560      \\
\end{tabular}
\end{ruledtabular}
\end{table}

Next, we test whether the Coulomb gauge calculation is working properly
by checking whether $I_n$ calculated in Coulomb gauge matches $I_n$
calculated in Feynman gauge for several choices of $n$. (The results
for Feynman gauge were checked against the program of Kunszt and Nason
\cite{KN} in Ref.~\cite{beowulfPRD}.) The results are presented in
Table \ref{tab:coulombvsfeynman}. We see that the results are properly
gauge invariant within the errors of the program.

\begin{table}[htb]
\caption{\label{tab:coulombvsfeynman}
Comparison of results in Feynman gauge and Coulomb
gauge for moments $I_n$ of the thrust distribution, Eq.~(\ref{thrustdef}). The
first error is statistical, the second systematic (determined from the
sensitivity to certain cutoffs used to control roundoff errors). We
choose
$\mu =\sqrt s$.}
\begin{ruledtabular}
\begin{tabular}{lll}
\multicolumn{1}{c}{n}
&\multicolumn{1}{c}{Feynman gauge}
&\multicolumn{1}{c}{Coulomb gauge}\\
\colrule
1.5 & $\ 4.127 \pm  0.008 \pm 0.025$ 
    & $\ 4.118 \pm  0.010 \pm 0.020$ \\
2.0 & $\ 1.565 \pm  0.002 \pm 0.007$ 
    & $\ 1.561 \pm  0.003 \pm 0.006$ \\
2.5 & $(6.439 \pm  0.010 \pm 0.022)\times 10^{-1}$ 
    & $(6.423 \pm  0.013 \pm 0.021)\times 10^{-1}$ \\
3.0 & $(2.822 \pm  0.005 \pm 0.009)\times 10^{-1}$ 
    & $(2.816 \pm  0.006 \pm 0.009)\times 10^{-1}$ \\
3.5 & $(1.296 \pm  0.002 \pm 0.004)\times 10^{-1}$ 
    & $(1.294 \pm  0.003 \pm 0.004)\times 10^{-1}$ \\
4.0 & $(6.159 \pm  0.011 \pm 0.016)\times 10^{-2}$
    & $(6.156 \pm  0.015 \pm 0.018)\times 10^{-2}$ \\
4.5 & $(3.009 \pm  0.006 \pm 0.007)\times 10^{-2}$
    & $(3.010 \pm  0.008 \pm 0.009)\times 10^{-2}$\\
5.0 & $(1.501 \pm  0.003 \pm 0.003)\times 10^{-2}$
    & $(1.503 \pm  0.004 \pm 0.004)\times 10^{-2}$\\
\end{tabular}
\end{ruledtabular}
\end{table}

\begin{figure}
\includegraphics[width = 8 cm]{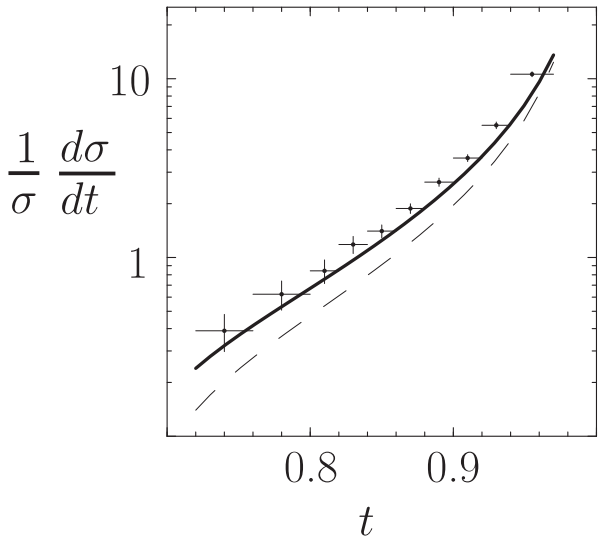}
\medskip
\caption{The thrust distribution at $\sqrt S = M_Z$ calculated in
Coulomb gauge at next-to-leading order. We also show, with a dashed
curve, the same distribution calculated at leading order. In both
cases, the renormalization scale is chosen to be $\mu = \sqrt S/6$. We
take $\alpha_s(M_Z) = 1.118$. The difference between the two theory
curves can be taken as an indication of the theory error arising from
neglect of graphs beyond order $\alpha_s^2$. The theory curves are
compared to data from the Opal Collaboration \protect\cite{thedata}.}
\label{fig:thrustdist}
\end{figure}

Having seen that the program appears to be working properly, we exhibit
a graph of the next-to-leading order thrust distribution ${\cal
I}(t)$ versus t in Fig.~\ref{fig:thrustdist}. We also show the same
distribution calculated at leading order and data from the
Opal Collaboration \cite{thedata}. The theoretical results are rather
sensitive to the choice of the $\overline{\rm MS}$ renormalization
scale $\mu$. We have chosen $\mu$ to be half of a typical jet energy in
a three jet event, $\sqrt s /3$. That is, $\mu = \sqrt s /6$. The
agreement between theory and data is not perfect, but this is to be
expected in a strictly perturbative expansion that includes only the
first two terms and no correction for effects beyond perturbation
theory such as hadronization effects.

As mentioned in the introduction,  one may wish to go beyond pure
next-to-leading order calculations by incorporating, in an approximate
way, some effects at all orders in $\alpha_s$. For instance, one may
want to  include renormalon effects by letting $\alpha_s$ run as a
function of loop momenta inside graphs. Alternatively, one may also
want to simulate realistic final states by adding parton showers to
the next-to-leading order calculation.  For such applications, one
must approximate, and the presence of unphysical degrees of freedom
propagating over long distances makes approximation difficult. A
straightforward  remedy is to do the calculation in a physical gauge,
such as Coulomb gauge. In this paper we have seen how this goal can be
accomplished.

\appendix

\section{Results for Feynman gauge}
\label{sec:appendix}

It is useful to have at hand the formulas for Feynman gauge that are
analogous to what we have found in Coulomb gauge. We present the needed
formulas in this appendix.

For the gluon propagator with a self-energy insertion, there is a
problem that we treat as described in \cite{beowulfPRL,beowulfPRD}.
The problem is most easily seen with the virtual self-energy graph. We
know that $\Pi^{\alpha\beta} \propto (q^2 g^{\alpha\beta} - q^\alpha
q^\beta)$. The first term is fine since it contains a factor $q^2$ that
cancels the $1/q^2$ in the adjoining propagator. The next term does not
have this good property, but its contribution will cancel in a sum over
graphs because it is proportional to $q^\alpha$ and $q^\beta$. In order
to make this cancellation happen in a single graph, we replace
\begin{equation}
(-g^\mu_\alpha)(-g^\nu_\beta)\Pi^{\alpha\beta}
\end{equation}
by
\begin{equation}
(-g^\mu_\alpha + q^\mu\tilde q_\alpha/\tilde q^2)
(-g^\nu_\beta + q^\nu\tilde q_\beta/\tilde q^2)
\Pi^{\alpha\beta}.
\end{equation}
This does not change the answer after summing over graphs because the
added terms are proportional to $q^\mu$ or $q^\nu$. But now the 
$q^\alpha q^\beta$ term in $\Pi^{\alpha\beta}$ gives zero in each graph
because $(-g^\mu_\alpha + q^\mu\tilde q_\alpha/\tilde q^2)q^\alpha = 0$.
We make this replacement for both the virtual and real versions of
$\Pi^{\alpha\beta}$.

For the real gluon self-energy graph, Eq.~(\ref{calMcoefs}) becomes
\begin{eqnarray}
N_{TT} &=&
2 C_A\bigl\{
-1 + x(1-x)
\bigr\}
+ N_F\bigl\{
1 - 2 x(1-x)
\bigr\},
\nonumber\\
N_{tt} &=&
4 C_A \,
x(1-x)  
- 4 N_F\, x(1-x),
\nonumber\\
N_{EE} &=&
-C_A\bigl\{
1+4x(1-x)
\bigr\}
+ 4 N_F\, x(1-x),
\nonumber\\
N_{Et} &=&
-2 C_A (2x-1)
+2 N_F (2x-1).
\label{calMcoefsFeynGauge}
\end{eqnarray}

For the virtual gluon self-energy graph, Eq.~(\ref{ATcoefs0}) becomes
\begin{eqnarray}
A^{\prime\prime}_{T,0}&=&
C_A{ 1 \over q^2}\biggl\{
{ 1 \over 2}\,\bar q^2
- { 5 \over 2}\, q^2 
+ 2 \bar q^2\,x(1-x)
\biggr\}
+N_F{ 1 \over q^2}\biggl\{
\bar q^2 
-{x(1-x)\over (1-\epsilon)}\ 2\bar q^2
\biggr\},
\nonumber\\
A^{\prime\prime}_{T,1}&=&
0,
\nonumber\\
A^{\prime\prime}_{T,2}&=&
0.
\label{ATcoefs0FeynGauge}
\end{eqnarray}
After subtracting $q^2 A_T(q^2)$ at $q^2 = 0$ from $q^2 A_T(q^2)$, we
are left with the revised version of Eq.~(\ref{ATcoefs1}),
\begin{eqnarray}
A^{\prime}_{T,0}&=&
2\,C_A \bigl\{ -1 + x(1-x)
\bigr\}
+ N_F 
\biggl\{
1
-{2 x(1-x)\over (1-\epsilon)}
 \biggr\},
\nonumber\\
A^{\prime}_{T,1}&=&
0 ,
\nonumber\\
A^{\prime}_{T,2}&=&
0 .
\label{ATcoefs1FeynGauge}
\end{eqnarray}
After renormalization, we have the revised version of
Eq.~(\ref{ATcoefs}),
\begin{eqnarray}
A_{T,0}&=&
-(2 C_A - N_F)\,
{ q^2 + e^2\mu^2 \over \bar q^2 + e^2\mu^2}\,
+ 2 C_A \, x(1-x)\,
{ q^2 + e^{5/3}\mu^2 \over \bar q^2 + e^{5/3}\mu^2}
\nonumber\\
&&   - 2 N_F\, x(1-x)\,
{ q^2 + e^{8/3}\mu^2 \over \bar q^2 + e^{8/3}\mu^2},
\nonumber\\
A_{T,1}&=&0,
\nonumber\\
A_{T,2}&=&0 .
\label{ATcoefsFeynGauge}
\end{eqnarray}

In Feynman gauge, the second term in the decomposition
Eq.~(\ref{Fgtensor0}) is not there, so that
\begin{equation}
F_g^{\mu\nu}(q)=
D(q)^{\mu\nu}
A_T(q^2).
\label{Fgtensor0FeynGauge}
\end{equation}
That is, $A_E(q^2) = A_T(q^2)$ in Eq.~(\ref{Fgtensor0}).

For the quark propagator with a self-energy insertion, we simply change
the gauge in the previous Coulomb gauge calculation. Then the
coefficients for a quark propagator with a cut self-energy diagram are
given by a revised form of Eq.~(\ref{calMqcoefs}),
\begin{eqnarray}
N_L &=&
C_F\bigl\{
 4 x(1-x) + (2x-1)(2x+\Delta)
\bigr\},
\nonumber\\
N_{E} &=&
2 C_F\,{ (1-x)},
\nonumber\\
N_{t} &=&
2C_F.
\label{calMqcoefsFeynGauge}
\end{eqnarray}
For the virtual quark self-energy, we begin with the revised form of 
Eq.~(\ref{BLcoefs0}),
\begin{eqnarray}
B^{\prime\prime}_{L,0}&=&
C_F
(1 - \epsilon),
\nonumber\\
B^{\prime\prime}_{L,1}&=&0,
\nonumber\\
B^{\prime\prime}_{L,2}&=&
0.
\label{BLcoefs0FeynGauge}
\end{eqnarray}
Then $B_{L,J}^{\prime} = B_{L,J}^{\prime\prime}$. After
renormalization, we have the revised form of Eq.~(\ref{BLcoefs}),
\begin{eqnarray}
B_{L,0}&=&
C_F\,
{ q^2 + e^1\mu^2 \over \bar q^2 + e^1\mu^2}\ ,
\nonumber\\
B_{L,1}&=&
0,
\nonumber\\
B_{L,2}&=&
0.
\label{BLcoefsFeynGauge}
\end{eqnarray}

In Feynman gauge, the second term in the decomposition
(\ref{quarkpropform1}) is not there, so that
\begin{equation}
F_q(q) = \rlap{/}q\,B_L(q^2).
\label{quarkpropform1FeynGauge}
\end{equation}
That is, $B_E(q^2)=0$ in Eq.~(\ref{quarkpropform1}).

The coefficients needed in the renormalization of the
quark-antiquark-vector boson three point functions change when we go
from Coulomb gauge to Feynman gauge. Specifically, in place of
Eq.~(\ref{Gammacoefs1}) we have
\begin{eqnarray}
&&C= C_A/2,
\nonumber\\
&&A_1(\epsilon) = -2, \ \ \ \ \
A_2(\epsilon) = 0, \ \ \
A_3(\epsilon) = -4(1-\epsilon),
\nonumber\\
&& A_4(\epsilon) = 0, \ \ \ \ \ \ \
A_5(\epsilon) = 0, \ \ \ 
A_6(\epsilon) = 0,
\nonumber\\
&& B_1(\epsilon) = 2, \ \ \ \ \ \ \
B_2(\epsilon) = 0.
\label{Gammacoefs1FeynGauge}
\end{eqnarray}
In place of Eq.~(\ref{Gammacoefs2}) we have
\begin{eqnarray}
&&C = C_F - C_A/2 = -1/(2N_C),
\nonumber\\
&&A_1(\epsilon) = -2(1-\epsilon), \ \
A_2(\epsilon) = 0, \ \
A_3(\epsilon) = 4(1-\epsilon),
\nonumber\\
&&A_4(\epsilon) = 0, \ \ \ \ \ \ \ \ \ \ \ \ \
A_5(\epsilon) = 0, \ \ 
A_6(\epsilon) = 0,
\nonumber\\
&& B_1(\epsilon) = 2, \ \ \ \ \ \ \ \ \ \ \ \ \ 
B_2(\epsilon) = 0.
\label{Gammacoefs2FeynGauge}
\end{eqnarray}

\begin{acknowledgments}
This work was supported in part by the U.S.~Department of Energy,
by the European Union under contract HPRN-CT-2000-00149, and by
the British Particle Physics and Astronomy Research Council.
\end{acknowledgments}

%23456789012345678901234567890123456789012345678901234567890123456789012
\end{document}